\documentclass[10pt, final]{IEEEtran}

\usepackage{lipsum}
\usepackage{amsthm}
\usepackage{amsmath}
\usepackage{bbm}

\usepackage{algorithm}
\usepackage{algorithmic}

\usepackage{stfloats}

\usepackage{mathrsfs}
\usepackage{cite}
\usepackage{bm,epsfig,amsthm,url}
\usepackage{indentfirst}
\usepackage{amssymb}
\usepackage{amsfonts}
\usepackage{epstopdf}
\usepackage[caption=false,font=footnotesize]{subfig}
\usepackage{xcolor}
\usepackage{mathtools}

\usepackage{hyperref}
\usepackage{balance} 

\theoremstyle{definition}
\newtheorem{lemma}{Lemma}
\newtheorem{theorem}{Theorem}

\allowdisplaybreaks[4]

\usepackage{graphicx}
\usepackage{epstopdf}

\usepackage{framed}

\begin{document}
	
\title{Multi-Functional Beamforming Design for Integrated Sensing, Communication, and Computation} 
	
\author{
Yapeng~Zhao, Qingqing~Wu,
~Wen~Chen,
~Yong~Zeng,
Ruiqi~Liu,
~Weidong~Mei,
~Fen~Hou,
~and~Shaodan~Ma

\thanks{Y. Zhao is with the State Key Laboratory of Internet of Things for Smart City, University of Macau, Macao 999078, China, and also with the Department of Electronic Engineering, Shanghai Jiao Tong University, 200240, China (email: zhao.yapeng@connect.um.edu.mo).  	
Q. Wu and W. Chen are with the Department of Electronic Engineering, Shanghai Jiao Tong University, 200240, China (e-mail: qingqingwu@sjtu.edu.cn; wenchen@sjtu.edu.cn). 
Y. Zeng is with the National Mobile
Communications Research Laboratory and Frontiers Science Center for Mobile Information Communication and Security, Southeast University, Nanjing
210096, China, and also with the Purple Mountain Laboratories,
Nanjing 211111, China (e-mail: yong\textunderscore zeng@seu.edu.cn). 
R. Liu is with the State Key Laboratory of Mobile Network and Mobile Multimedia Technology, ZTE Corporation, Shenzhen 518057, China (e-mail: richie.leo@zte.com.cn).
W. Mei is with the National Key Laboratory of Wireless Communications,
University of Electronic Science and Technology of China, Chengdu 611731,
China (e-mail: wmei@uestc.edu.cn).
F. Hou and S. Ma are with the State Key Laboratory of Internet of Things for Smart City, University of Macau, Macao 999078, China (email: fenhou@um.edu.mo; shaodanma@um.edu.mo).}

}

\maketitle
	
\begin{abstract} 
Integrated sensing and communication (ISAC) systems may face a heavy computation burden since the sensory data needs to be further processed.
This paper studies a novel system that integrates sensing, communication, and computation, aiming to provide services for different objectives efficiently. 
This system consists of a multi-antenna multi-functional base station (BS), an edge server, a target, and multiple single-antenna communication users. The BS needs to allocate the available resources to efficiently provide sensing, communication, and computation services. Due to the heavy service burden and limited power budget, the BS can partially offload the tasks to the nearby edge server instead of computing them locally. 
We consider the estimation of the target response matrix, a general problem in radar sensing, and utilize Cram\'{e}r-Rao bound (CRB) as the corresponding performance metric. 
To tackle the non-convex optimization problem, we propose both semidefinite relaxation (SDR)-based alternating optimization and SDR-based successive convex approximation (SCA) algorithms to minimize the CRB of radar sensing while meeting the requirement of communication users and the need for task computing. 
Furthermore, we demonstrate that the optimal rank-one solutions of both the alternating and SCA algorithms can be directly obtained via the solver or further constructed even when dealing with multiple functionalities.
Simulation results show that the proposed algorithms can provide higher target estimation performance than state-of-the-art benchmarks while satisfying the communication and computation constraints.

\end{abstract}

\section{Introduction}
It has been visioned for years and gradually reached a consensus that the upcoming sixth-generation (6G) network will be a multi-functional network which aims to integrate communication, sensing, computing, and intelligence \cite{WalidSaad_6GVision, YuanmingShi_JSAC22_EdgeAI, ZhongxiangWei_ComMag22_MultiFunctional, LiuLLCLS23}. 
Such integration, acting as the powerful engine for realizing the intelligent world of the future, will play a crucial role in providing efficient services for computation-intensive, communication-intensive, and delay-sensitive applications, e.g., smart factories and autonomous driving  \cite{YuanmingShi_Survey20_EdgeAI, JSTSP23_IRS_Computing}. 
Therefore, there has been a noticeable surge of interest in the exploration of how to efficiently realize integrated sensing and communication (ISAC) \cite{FanLiu_JSAC22, li2023towards, ISAC_ProcIEEE11}, integrated communication and computation \cite{FL_IRS23, AirComp_survey}, further the multiple functionalities in integrated sensing, communication, and computation (ISCC) systems  \cite{XiaoyangLi_Integrated_SensingComputation, GuangxuZhu_Nwtwork23_ISCC, LindongZhao_TWC22_ISCC, YinghuiHe_TWC24_ISCC}.
	
Recent works have demonstrated the potential to commercialize ISAC in the future 6G network since the wireless sensing and communication systems are evolving more similarly.
For example, they are both moving to higher frequency, large-scale antenna arrays as well as hardware architectures and signal processing techniques \cite{RangLiu_TWC24_RISISAC, XianxinSong_TWC24_IRS_ISAC, YuanhaoCui_Netw21_ISAC}. 
The ISAC system utilizes the shared wireless signal transmitter to simultaneously realize radar and communication functionalities \cite{HaochengHua_TWC24_MIMOISAC,10286534}.
The transmitter can allocate different weights to the radar and communication components, thereby constructing beams toward the targets and communication destinations \cite{li2023towards}, ultimately enabling flexible control over the beamforming process and ensuring efficient performance for both functionalities.
 
	
The sensory data in the ISAC system can be collected and further utilized to enhance various services, such as communication and localization. 
For instance, in sensing-aided vehicular beamforming \cite{KaitaoMeng_TVT23_VehicularISAC}, the sensory data collected from the environment can be leveraged to optimize beamforming techniques specifically tailored for vehicular communication scenarios. 
The inherent information in the obtained data needs to be fully excavated for target recognition, industrial control, autonomous driving \cite{KaitaoMeng_TVT23_VehicularISAC, YuanmingShi_Survey20_EdgeAI}, etc. 
Generally, there are three main kinds of process requirements in the ISAC system which facilitate the evolution of ISCC system. The first one is the requirement for direct sensing services, e.g., target recognition, tracking, and predicting target movement. Specifically, the raw data collected by the system may need to be quantized and conveyed to the neural network for target recognition. For tracking and prediction, the spatiotemporal characteristics of the echo signal need to be analyzed.
The second type is the requirement for upper-level services that are based on the sensing results and take them a step further, such as simultaneous localization and mapping \cite{MengHua_WCL23_IRS_Localization, KaitaoMeng_TCOM_Localization}, vehicle platooning, etc. Besides, in practice, various types of sensors that can provide unique information can be harnessed in addition to radio-sensed data, including cameras, lidar, and other sensing modalities. The multimodal data fusion can enhance the perception capabilities of the vehicles but bring huge computation requirements challenges.
Thirdly, the base station (BS) and devices may have additional computation tasks that need to be efficiently processed \cite{GuangjiChen_DIBF_MEC}. For example, the devices may have delay-sensitive tasks that require to be efficiently executed, and the data collected by the BS needs to be further processed, e.g. cooperatively training for further target recognition \cite{NingHuang_IoT23_UAV_ISCC}. 
All the three aforementioned computation requirements involve complex signal processing and data analysis, which usually necessitates substantial computational resources. The ISAC system may have a heavy computing burden as mentioned above. However, it cannot process data timely due to the limited power/energy budget and computing capability of the participating nodes. Besides, the BS must handle multiple responsibilities including communication and sensing tasks. 
Hence, the ISCC system, which aims to coordinate the available resources and further provide efficient computing services for practical ISAC systems, has attracted wide attention.

How to integrate computing functionality into ISAC systems to form ISCC is still unclear, but we can learn lessons from the existing works on ISAC. 
In current literature related to ISAC and radar sensing, various performance metrics were considered. For instance, the estimated mutual information \cite{YifengXiong_TIT23_MutualInformation}, signal-to-interference-plus-noise ratio (SINR) of reflected echo \cite{MengHua_TWC23_SecureIRS_ISAC, KaitaoMeng_TWC23_UAV_ISAC}, transmit beampattern error compared to the desired one \cite{XiangLiu_TSP20_MIMORadar}, and Cram\'{e}r-Rao bound (CRB) \cite{YongZeng_TSP24_NF_CRB, peng2024semi} that provides the lower bound of variance for any unbiased estimators.
The beampattern gain at the target was maximized in \cite{MengHua_TWC23_SecureIRS_ISAC} while ensuring the communication service. The tradeoff between CRB and rate in ISAC has been studied in the context of multiple single-antenna communication users \cite{FanLiu_TSP22_CRB} and multicast communication \cite{ZixiangRen_TWC24_ISAC_multicast}, respectively.
Furthermore, both the traditional communication-centric and novel sensing-centric energy efficiency maximization problem of ISAC was considered in \cite{JiaqiZou_TCOM24_ISAC_EE}.
With the growing demand for computing applications in future networks, there have emerged some works focused on combining ISAC with mobile edge computing (MEC) techniques  \cite{YinghuiHe_TWC24_ISCC, MinghuiDai_TITS21}. Such a system involves sensing, communication, and computation functionalities, and how to efficiently allocate the available resources to support the system is challenging.	Specifically, the sensing accuracy was maximized in \cite{YinghuiHe_TWC24_ISCC} while meeting the delay constraints of the tasks. Furthermore, the corresponding real-world test was conducted to validate the proposed ISCC system. In \cite{Ding_JSAC22_ISCC, NingHuang_IoT23_UAV_ISCC, BinLi_UAV_ISCC_TGCN23}, the dual-function radar communication devices were assumed to perform radar sensing tasks and upload collected data to the BS simultaneously. In particular, the weighted minimization problem of energy consumption of computation and square error of radar beampattern was studied in \cite{Ding_JSAC22_ISCC}. Additionally, \cite{BinLi_UAV_ISCC_TGCN23} investigated the impact induced by the association between users and unmanned aerial vehicles (UAVs), as well as UAV trajectory design. 
In \cite{PengLiu_ISCC}, the task execution time was minimized while satisfying the SINR requirement for communication and sensing.

There still remain some gaps in the existing works on ISCC that need to be addressed.
Firstly, previous works primarily focus on implicit requirements related to radio sensing, e.g., the SINR of the received echo signal or the transmit beampattern. However, the CRB, regarded as a fundamental and explicit estimation accuracy limit for a given sensing scenario \cite{FanLiu_TSP22_CRB}, has yet to be deeply analyzed in the ISCC system. 
Note that incorporating CRB in ISCC design could lead to improved sensing capabilities and decision-making processes.
Second, existing works on ISCC tend to neglect the communication needs of devices and only focus on the sensing and task offloading processes. However, considering the communication requirements of devices is crucial for designing efficient ISCC systems. Communication plays a pivotal role in facilitating crucial aspects such as data exchange, synchronization, coordination, and cooperative decision-making among devices within the ISCC system. Neglecting the communication aspect can lead to impractical systems that fail to ensure efficient and effective data exchange and collaboration among devices.
	
To address these issues, we propose a design framework for the ISCC system in this paper, which includes one multi-functional BS, multiple communication users, one edge server, and one target. Specifically, the BS needs to provide the communication and computation services, and meanwhile sense the target via the transmitted radio waveform.
We minimize the corresponding achieved CRB under both point and extended target scenarios, while meeting the communication and computation constraints as well as the services requirement.
However, the performance of each function is affected by others since the data collection, transmission, and computation are coupled together. How to schedule the resources and balance the performance among the triple functions is a major challenge.
We aim to minimize the achievable CRB, subject to the SINR constraints for the communication users, the computation requirement, and the power budget on the BS. 
Through this problem, we show that a flexible trade-off among sensing, communication, and computation can be achieved.
The main contributions can be summarized as follows.
	\begin{itemize}
		\item The multi-functional beamforming for the ISCC system is capable of providing the three services efficiently via controlling the power allocation and transmit beam direction. We propose a general design framework that concentrates on minimizing the achievable CRB for both point target and extended target scenarios by jointly optimizing the transmit beampattern and local computing frequency, subject to individual communication SINR constraints, computation requirement, and the transmit power budget. This framework can achieve a flexible tradeoff among sensing, communication, and computation functionalities by adjusting the corresponding constraints.
		\item A significant challenge in this problem lies in the precise allocation of power to different functionalities since the transmit beamforming needs to mitigate the interference among communication users and the edge server. The non-convex property of the optimization problem arises due to two primary aspects: the need to simultaneously realize multiple functionalities which is competitive and the coupled optimization variables. We first analyze the inherent characteristics of this system to gain useful insights and thus simplify the original formulation. Next, we introduce auxiliary variables to further convert it to an equivalent but more tractable form. Ultimately, we propose the semidefinite relaxation (SDR)-based alternating optimization (AO) algorithm, and additionally, the SDR-based successive convex approximation (SCA) algorithm that can iteratively converge to better performance.
		\item The solutions obtained by semidefinite programming (SDP) may have a higher rank than one, in that case, additional steps are needed to extract feasible but suboptimal solutions. However, this procedure may induce inevitable performance loss. To avoid this drawback and find the exact global optimum for both the two problems, we analyze and further demonstrate that the obtained solutions for both problems are rank-one in general or can be constructed without loss the optimality. 
		Moreover, we provide insights for the special case that there is no communication requirement, which is beneficial for engineering design. 
		\item Simulation results verified that the proposed algorithms are capable of efficiently accommodating communication, computation, and sensing functionalities by precisely allocating the available resources. The proposed algorithms can efficiently construct multi-functional transmit beamforming to achieve the tradeoff between these objectives while avoiding the performance loss associated with the SDR technique.
		 It is observed that both the SDR-based AO and SDR-based SCA algorithms can achieve lower CRB compared to the traditional beampattern approximation approach. 
		 Besides, the proposed ISCC can enhance a particular function at a small fraction of the performance loss of other services.
		Overall, the numerical results demonstrate that the ISCC system can simultaneously realize and flexibly accommodate these three functionalities via precise power allocation and beamforming design.
	\end{itemize}
	
\emph{Notation:} $\mathbb{E}[\cdot] $ denotes the statistical expectation. 
Besides, ${\cal O}\left(  \cdot  \right)$ represents the big-O computational complexity notation.
Boldface lowercase and uppercase letters denote the vectors and matrices, respectively, while scalars are denoted by italic letters. The space of $m\times n$ real-valued matrices is denoted by $\mathbb{R}^{m\times n}$, and the space of complex-valued matrices is denoted by $\mathbb{C}^{m\times n}$.
For any complex-valued vector $\mathbf{x}$, ${\left\| \mathbf{x} \right\|}$ denote the corresponding Euclidean norm. 
Regarding a matrix $\mathbf{A}$, the notation $\mathbf{A}^{H}$ represents its conjugate transpose, $\mathrm{rank}(\mathbf{A})$ denotes its rank, $\mathrm{tr}(\mathbf{A})$ represents its trace, and $\mathbf{A}_{i,j}$ denotes the entry at the $i$th row and $j$th column, respectively. 
The notation ${\mathbf{A}} \succeq {\mathbf{0}}$ indicates that the matrix $\mathbf{A}$ is positive semi-definite. 

\section{System Model and Problem Formulations}
	
\begin{figure}[t]  
	\centering
	\includegraphics[width=0.8\columnwidth]{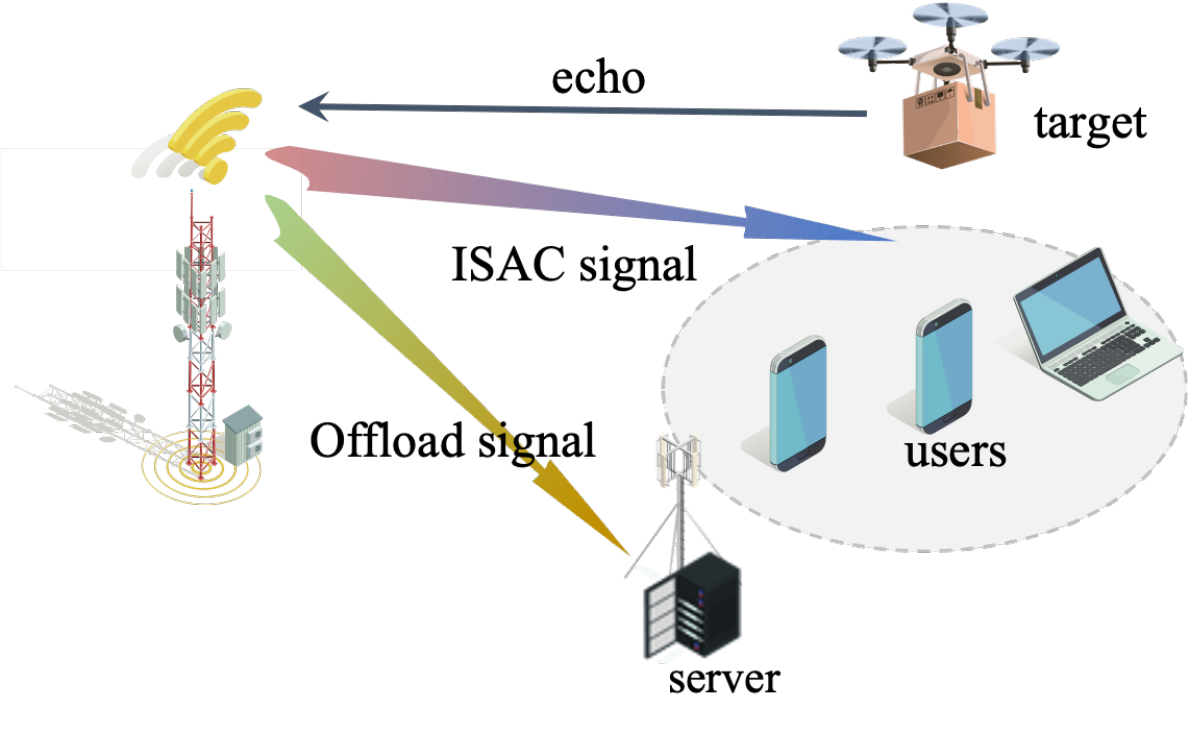} 
	\caption{The ISCC system model.}
	\label{fig_model}
\end{figure}
	
We assume that a multi-functional BS aims to provide sensing, communication, and computation tasks for the whole network as depicted in Fig. \ref{fig_model}. Specifically, the BS serves $K$ single-antenna communication users and optionally offloads the tasks to the single-antenna edge server for fast computing. Besides, the transmission signal wave is exploited to sense the potential targets simultaneously, e.g., the invading UAV.
We assume that the target is near the BS, and the corresponding radio propagation is line-of-sight. Besides, the channels from the BS to the users and the server follow the Rician distribution. These channels remain static during the considered ISAC period which spans $T$ symbols.
The BS is equipped with $M = M_t + M_r$ antennas, where $M_t$ antennas are for transmitting the ISAC signal, and the remaining $M_r$ antennas are left for receiving the radar echo. The transmit signal vector in $t$-th slot, $t \in \{1, 2, \dots, T\}$ is 
\begin{align}
	\mathbf{s}[t] = \sum_{k=1}^{K+1} \mathbf{w}_{c,k} x_{c,k}[t] +  \mathbf{W}_{r} \mathbf{x}_{r}[t],
\end{align}
where the radar signal is denoted as $x_{r,m}[t]$ and satisfies $ \mathbb{E} \{x_{r,m}[t]\} = 0, \mathbb{E} \{x_{r,m}[t]x_{r,m}^*[t]\} = 1 $, and $\mathbf{w}_{r,m} \in \mathbb{C}^{M_t \times 1}$ denotes the corresponding radar beamfomer.
We denote $\mathbf{W}_{r} = [\mathbf{w}_{r,1}, \cdots, \mathbf{w}_{r,M_t}]$, $\mathbf{x}_{r}[t] = [x_{r,1}[t]; \ldots; x_{r,M_t}[t]]$. 
Besides, $x_{c,k}[t] \in \mathbb{C}, \forall 1\leq k\leq K$, and $x_{c,K+1}[t] \in \mathbb{C}$ denote the transmit signal for user $k$ and the server, respectively. The transmit data symbols satisfy $x_{c,k}[t] \sim \mathcal{CN}(0,1), 1\leq k \leq K+1$. Besides, the corresponding transmit beamformer for communication is denoted by $\mathbf{w}_{c,k}\in \mathbb{C}^{M_t\times 1}$. 
Without loss of generality, the communication and radar signals are assumed to be statistically independent. Hence, the power consumption due to the signal transmission is given by 
\begin{align}
	\sum_{k=1}^{K+1} \|\mathbf{w}_{c,k}\|^2 + \|\mathbf{W}_{r}\|^2_F.
\end{align}

\subsection{Sensing Model}
Let $\mathbf{S}=[\mathbf{s}[1],\dots, \mathbf{s}[T]]$ denote the transmitted signal over the $T$ symbols. 
It is assumed that 
\begin{align}
	\frac{1}{T} \mathbf{S} \mathbf{S}^H \approx \mathbf{R}_s =  \sum_{k=1}^{K+1}\mathbf{w}_{c,k}\mathbf{w}_{c,k}^H  + \mathbf{W}_{r} \mathbf{W}_{r}^H,
\end{align}
since $T$ is sufficiently large. 
We assume that the BS can suppress the self-interference sufficiently via the corresponding techniques \cite{SabharwalSGBRW14a}.
The received echo signal is
\begin{align}
	\mathbf{Y} = \mathbf{G} \mathbf{S} + \mathbf{N},
\end{align}
where $\mathbf{G}\in \mathbb{C}^{M_r \times M_t}$ is the target response matrix (TRM) and $\mathbf{N} \in \mathbb{C}^{M_r \times T} $ denotes the additive white Gaussian noise (AWGN) where each entry is random with independent and identically distributed (i.i.d.) with zero mean and variance $\sigma^2_s$. 
	
For the point target, the TRM is given by
\begin{align}
	\mathbf{G} = \alpha \mathbf{A}(\theta) = \alpha \mathbf{b}(\theta)\mathbf{a}^H(\theta),
\end{align}
where $\alpha$ and $\theta$ are the complex reflection coefficient and the angle of the target, $\mathbf{a}(\theta) \in \mathbb{C}^{M_t \times 1} $, $\mathbf{b}(\theta) \in \mathbb{C}^{M_r \times 1}$ are the transmitting and receiving steering vectors, respectively. 
For the uniform linear arrays (ULA) antennas, by choosing the first element as the antenna element, we have
\begin{align}
	&\mathbf{a}(\theta) = [1, e^{j 2\pi \Delta \sin \theta}, \dots, e^{j 2\pi (M_t -1) \Delta \sin \theta }]^T, \\
	&\mathbf{b}(\theta) = [1, e^{j 2\pi \Delta \sin \theta}, \dots, e^{j 2\pi (M_r -1) \Delta \sin \theta }]^T,
\end{align}
where $\Delta$ is the spacing between adjacent elements normalized by wavelength. We assume that the transmit and receive antennas are ULAs with half-wavelength antenna spacing.
The corresponding CRB for estimating $\theta$ is given by \cite{FanLiu_TSP22_CRB},
\begin{align} \label{CRB1}
		\!\! \mathrm{CRB}_1(\theta) \! =  \!\frac{ \frac{\sigma_s^2}{2|\alpha|^2 T} \mathrm{tr}(\mathbf{A}^H \mathbf{A} \mathbf{R}_s )  }{ \mathrm{tr}( \dot{\mathbf{A}}^H \dot{\mathbf{A}} \mathbf{R}_s) \mathrm{tr}(\mathbf{A}^H \mathbf{A} \mathbf{R}_s ) \! - \! |\mathrm{tr}( \dot{\mathbf{A}}^H \mathbf{A} \mathbf{R}_s)|^2},
	\end{align}
where 
	\begin{align}
		&\dot{\mathbf{A}} \triangleq \dot{\mathbf{A}}(\theta) = \frac{\partial \mathbf{A}(\theta)}{\partial \theta} =  \mathbf{b}(\theta) \dot{\mathbf{a}}^H(\theta) + \dot{\mathbf{b}}(\theta)\mathbf{a}^H(\theta), \\
		&\hspace{-8pt} \dot{\mathbf{a}} \triangleq \dot{\mathbf{a}}(\theta) = \frac{\partial \mathbf{a}(\theta)}{\partial \theta} = [1, \dots, j 2\pi a_{M_t} (M_t-1) \Delta \cos \theta  ]^T, \\
		&\hspace{-8pt} \dot{\mathbf{b}} \triangleq \dot{\mathbf{b}}(\theta) \! = \! \frac{\partial \mathbf{b}(\theta)}{\partial \theta} = [1, \dots, j 2\pi b_{M_r} (M_r-1) \Delta \cos \theta ]^T.
	\end{align}
The interpretation of minimizing the $\mathrm{CRB}_1(\theta)$ is to illuminate a particular direction with the azimuth angle of $\theta$ via the optimized transmit beampattern. 
	
For the extended target, the TRM is given by
	\begin{align}
		\mathbf{G} = \sum_{i=1}^{N_s} \alpha_i \mathbf{A}(\theta_i) = \sum_{i=1}^{N_s} \alpha_i \mathbf{b}(\theta_i)\mathbf{a}^H(\theta_i),
	\end{align}
where $N_s$ denotes the number of scatterers, $\alpha_i$ and $\theta_i$ denotes the reflection coefficient and the angle of the $i$-th scatterer, respectively. 
We choose the whole TRM $\mathbf{G}$ as a parameter to be estimated in the extended target scenario as \cite{FanLiu_TSP22_CRB, chen2022_ISAC_SWIPT}, and the corresponding CRB that adopts the trace of the CRB matrix as a scalar performance metric is given by
	\begin{align} \label{CRB2}
		\mathrm{CRB}_2(\mathbf{G}) = \frac{\sigma^2_s M_r}{T} \mathrm{tr}(\mathbf{R}_s^{-1}).
	\end{align}
Note that the dedicated radar stream without carrying communication data is necessary when estimating the entire TRM $\mathbf{G}$, i.e., the extended target case since the transmitted signal matrix needs to be full rank of $M_t$ to provide more degrees of freedom to estimate $\mathbf{G}$ \cite{FanLiu_TSP22_CRB}.

\subsection{Communication Model}
	
Since dedicated radar beams contain no useful information, the radar signals cause interference to the communication users and the edge server. 
Let $\mathbf{h}_k \in \mathbb{C}^{M_t\times 1} $ denote the channel vector from the multi-functional BS to user $k$, and $\mathbf{h}_{K+1} \in \mathbb{C}^{M_t\times 1} $ denote the channel vector from the multi-functional BS to edge server. Then, the SINR of user $k$ is given by
\begin{align}
	\Gamma_k = \frac{ |\mathbf{h}_k^H \mathbf{w}_{c,k}|^2 }{\sum_{i=1, i\neq k}^{K+1} |\mathbf{h}_k^H \mathbf{w}_{c,i}|^2 + \|\mathbf{h}_k^H \mathbf{W}_{r}\|^2 + \sigma^2_c }.
\end{align}
Note that there is no need for an additional radar beamformer $\mathbf{W}_r$, i.e., $\mathbf{W}_{r} = \mathbf{0}$, to estimate the angle $\theta$ in point target case \cite{FanLiu_TSP22_CRB}.
The achievable communication rate for user $k$ is $R_k = \log_2(1+\Gamma_k)$.

\subsection{Computation Model}
In each symbol $t$, the SINR for task offloading from the BS to the server is shown as
\begin{align}
	\gamma_p = \frac{ |\mathbf{h}_{K+1}^H \mathbf{w}_{c,K+1}|^2 }{\sum_{i=1}^{K} |\mathbf{h}_{K+1}^H \mathbf{w}_{c,i}|^2 + \|\mathbf{h}_{K+1}^H \mathbf{W}_{r}\|^2 + \sigma^2_c}.
\end{align}
Besides, the local computation rate at the BS is defined as the number of bits processed per second, i.e., $f/\rho$, where $f$ denotes the current central processor unit (CPU) frequency assigned for task computing, and $\rho$ denotes the number of CPU cycles required for completing $1$-bit data. Note that the server has sufficient computing capability and the computed outcome is much smaller than the offloaded task, thus the corresponding latency can be neglected \cite{JSTSP23_IRS_Computing}.
In that case, the task process rate is shown as the local computing rate plus the task offloading rate:
\begin{align} \label{process_rate}
	R_p = \frac{f}{\rho} + B\log_2(1+\gamma_p).
\end{align}
Generally, the edge server is charged from the power grid, and we mainly focus on the power allocation at the BS.
The corresponding power budget, that constrains the transmission and local computing, at the multi-functional BS is given by 
\begin{align} \label{p_CRBmin_power}
	\sum_{k=1}^{K+1} \|\mathbf{w}_{c,k}\|^2 +  \|\mathbf{W}_{r}\|_F^2 + \kappa f^3 \leq P_0,
\end{align}
where $\kappa$ is the computational energy efficiency of the CPU chip \cite{GuangjiChen_DIBF_MEC}, and $P_0$ is the maximum available power at the BS.

\subsection{Problem Formulation}

The objective of this paper is to reveal the performance tradeoff among communication rate, computation rate, and the achieved CRB.
We optimize the transmit beamforming to enhance the sensing quality, i.e., minimize the achieved CRB subject to minimum communication SINR constraint and task process constraint. Mathematically, the corresponding problem is given by
\begin{subequations} \label{p_CRBmin}
	\begin{align}
		\mathop  {\min  }\limits_{\scriptstyle \{\mathbf{w}_{c,k}\}, \mathbf{W}_{r}, f
				\atop \scriptstyle } \ \ \
		& \mathrm{CRB} \label{p_CRBmin_obj} \\
		{\rm{s.t.}}\ \ \ \ \ \ \ 
		& 0 \leq f \leq f_{\max}, \label{p_CRBmin_f} \\
		& \Gamma_k \geq \gamma_{k}, \forall k,\label{p_CRBmin_C} \\
		& f/\rho + B\log_2(1+\gamma_p) \geq R_{p,\min}, \label{p_CRBmin_P}\\
		& \eqref{p_CRBmin_power},
	\end{align}
\end{subequations}
where $\gamma_{k}$ is the minimum communication SINR requirement, and $R_{p,\min}$ represents the minimum computation requirement at the BS. Besides, \eqref{p_CRBmin_f} denotes the CPU computing frequency constraint at the BS due to the limited resources. Note that problem \eqref{p_CRBmin} is non-convex since there exist coupled variables in objective function \eqref{p_CRBmin_obj} and constraints \eqref{p_CRBmin_C}, \eqref{p_CRBmin_P}.  Furthermore, we demonstrate that the original problem can be optimally tackled through SDP with dropped rank-one constraints in the following sections.

\section{Algorithms For Point Target Case}

\subsection{No Communication User Case}
\label{subsection_no_comm1}
	
First, we consider the special case that there is no communication user in this ISCC system. In this scenario, we have $\mathbf{w}_{k} = \mathbf{0}, \gamma_{k} = 0, \forall 1 \leq k \leq K$, the corresponding problem is shown as
\begin{subequations} \label{P_no_comm}
	\begin{align}
		\mathop  {\max  }\limits_{\scriptstyle \mathbf{w}_{K+1}, f
			\atop \scriptstyle }\ \ \
		& |\mathbf{a}^H \mathbf{w}_{K+1}|^2 \label{P_no_comm_obj} \\
			{\rm{s.t.}}\ \ \ \;
		& 0 \leq f \leq f_{\max}, \label{} \\
		& \|\mathbf{w}_{K+1}\|^2 + \kappa f^3 \leq P_0, \\
		& \eqref{p_CRBmin_f} \eqref{p_CRBmin_P}.
	\end{align}
\end{subequations}
Note that problem \eqref{P_no_comm} is still non-convex that cannot be directly solved.
From problem \eqref{P_no_comm}, we have
\begin{align}
	\mathbf{w}_{K+1} \in \mathrm{span} \{\mathbf{a}, \mathbf{h}_{K+1} \},
\end{align}
since the transmit beamformer $\mathbf{w}_{K+1} $ needs to maximize the objective function \eqref{P_no_comm_obj} while enlarging the received SINR at the edge server to satisfy the computation requirement \eqref{p_CRBmin_P}. 
	
\begin{lemma}
	With any given $\mathbf{w}_{K+1}$, we have $f=0$ if the full power transmission can satisfy the computing requirement, i.e., 
	\begin{align} 
		B\log_2(1+ \gamma_p) \geq R_{p,\min}.
	\end{align}
\end{lemma}
	
{\it{Proof:}} It can be proved by contradiction. The objective function value will decrease if any power is assigned to the local computation, i.e. $f>0$. $\hfill\blacksquare$
	
Based on Lemma 1, we can obtain the following theorem.

\begin{theorem}
If 
	\begin{align} \label{no_comm_c}
		\sqrt{P_0} |\mathbf{h}_{K+1}^H \mathbf{a}|^2 \geq M_t (2^{R_{p,\min}}-1) \sigma_c^2,
	\end{align}
the optimal solution for problem \eqref{P_no_comm} is given by
	\begin{align}
		\mathbf{w}_{K+1} = \sqrt{P_0} \frac{\mathbf{a}}{\|\mathbf{a}\|}, f = 0.
	\end{align}
\end{theorem}

{\it{Proof:}} The objective function value is maximized when $\mathbf{w}_{K+1} $ is aligned with $\mathbf{a}$. In that case, if the maximum power transmission can satisfy the task offloading requirement, all the power needs to be allocated to maximize the objective function value, i.e., $f=0$. $\hfill\blacksquare$

In general, we can utilize the SDR technique to obtain a satisfying solution, the corresponding transformation is given by
	\begin{subequations} \label{P_no_comm1}
		\begin{align}
			\mathop  {\max  }\limits_{\scriptstyle \mathbf{W}_{K+1}, f
				\atop \scriptstyle } 
			&  \mathrm{tr}(\mathbf{A} \mathbf{W}_{K+1}) \\
			&\hspace{-40pt} {\rm{s.t.}}\
			f/\rho + B\log_2\Big(1+ \frac{ \mathrm{tr}(\mathbf{Q}_{K+1} \mathbf{W}_{K+1}) }{ \sigma^2_c }\Big) \geq R_{p,\min}, \label{}\\
			& \mathrm{tr}(\mathbf{W}_{K+1}) + \kappa f^3 \leq P_0, \\
			& \mathbf{W}_{K+1} \succeq \mathbf{0}, \\
			& \mathrm{rank}(\mathbf{W}_{K+1}) = 1, \\
			& \eqref{p_CRBmin_f}.
		\end{align}
	\end{subequations}
Note that the problem \eqref{P_no_comm1} is a convex SDP problem when the rank-one constraint is dropped, hence, it can be efficiently solved via existing solvers, e.g., CVX \cite{cvx}. 
However, if the optimal solutions have a higher rank than one, additional steps are required to extract a suboptimal solution, which may result in an inevitable performance loss. Fortunately, we demonstrate in Theorem \ref{theorem_point_1} that the obtained optimal solutions are generally rank-one.

\begin{theorem} \label{theorem_point_1}
	There always exists an optimal solution $\mathbf{W}_{K+1}$ satisfying $\mathrm{rank}(\mathbf{W}_{K+1}) = 1$ for problem \eqref{P_no_comm1}.
\end{theorem}
{\it{Proof:}} See Appendix A. $\hfill\blacksquare$
	
Theorem \ref{theorem_point_1} shows that the obtained optimal solution always is rank-one, thus we can directly recover the transmit precoder $\mathbf{w}_{K+1}$ without loss of optimality.
In summary, when facing problem \eqref{P_no_comm}, we first check whether \eqref{no_comm_c} is satisfied under the full power transmission. If satisfied, we can obtain the corresponding closed-form solution, otherwise, we resort to the problem \eqref{P_no_comm1} to obtain the optimal solution and then recover the transmit precoder $\mathbf{w}_{K+1}$.

\subsection{SDR-based AO Approach}
	
To address the non-convexity issue for the general case that multiple devices exist in the ISCC system, a natural way is to reformulate it as an SDP problem by the matrix lifting technique. Specifically, by defining $\mathbf{W}_{c,k} = \mathbf{w}_{c,k} \mathbf{w}_{c,k}^H, \forall 1 \leq k \leq K+1$, we have $	|\mathbf{h}_{k}^H \mathbf{w}_{c,k}|^2 = \mathrm{tr}( \mathbf{Q}_k \mathbf{W}_{c,k})$. Then, by leveraging the Schur complement theory \cite{Schur_Book}, problem \eqref{p_CRBmin} for sensing the point target can be equivalently transformed to
		\begin{subequations} \label{P1}
		\begin{align}
			\!\!\!\!\!\! \mathop  {\min  }\limits_{\scriptstyle \{\mathbf{W}_{c,k}\},
				\atop  f, t \scriptstyle } \ \ \
			& -t \\
			{\rm{s.t.}}\ \ \ \;
			& \mathbf{W}_{c,k}, \mathbf{Z}(\mathbf{R}_s)\succeq \mathbf{0}, \forall 1\leq k \leq K+1, \label{p_SDR_postive} \\
			& \mathrm{rank}(\mathbf{W}_{c,k}) = 1, \forall 1\leq k \leq K+1, \label{p_SDR_rank_1} \\
			& \sum_{k=1}^{K+1} \mathrm{tr}( \mathbf{W}_{c,k})  + \kappa f^3 \leq P_{0}, \label{p_SDR_power} \\
			&\eqref{p_CRBmin_f}, \eqref{p_CRBmin_C},\eqref{p_CRBmin_P},
		\end{align}
	\end{subequations}
	where
	\begin{align}
		&\!\!\!\!\mathbf{Z}(\mathbf{R}_s) \triangleq \left[                
		\begin{array}{lll}   
			\mathrm{tr}(\dot{\mathbf{A}}^H \dot{\mathbf{A}} \mathbf{R}_s)-t    &   \mathrm{tr}(\dot{\mathbf{A}}^H \mathbf{A} \mathbf{R}_s ) \\  
			\mathrm{tr}(\mathbf{A}^H \dot{\mathbf{A}} \mathbf{R}_s ) &     \mathrm{tr}(\mathbf{A}^H \mathbf{A} \mathbf{R}_s) \\ 
		\end{array} \right] \nonumber \\
		&\!\!\!\!\!\! =\left[\begin{array}{cc} \!\!\!
			\|\dot{\mathbf{b}}\|^2 \mathbf{a}^H \mathbf{R}_X \mathbf{a}+\|\mathbf{b}\|^2 \dot{\mathbf{a}}^H \mathbf{R}_X \dot{\mathbf{a}}-t \!\!\! & \|\mathbf{b}\|^2 \mathbf{a}^H \mathbf{R}_X \dot{\mathbf{a}} \\
			\|\mathbf{b}\|^2 \dot{\mathbf{a}}^H \mathbf{R}_X \mathbf{a} & \|\mathbf{b}\|^2 \mathbf{a}^H \mathbf{R}_X \mathbf{a}
		\end{array}\right].
	\end{align}
It is observed that problem \eqref{P1} is still non-convex due to the rank-one constraint induced by the matrix-lifting process and the coupled variables in constraint \eqref{p_CRBmin_P}. To tackle this difficulty, we first divide the optimization variables into two distinct blocks and then reformulate the two subproblems into a more tractable form.
	
	The first optimization problem w.r.t.  variables $\mathbf{w}_{c,K+1},  f, t$ is shown as
	\begin{subequations} \label{P_AO1}
		\begin{align}
			\!\!\!\!\!\! \mathop  {\min  }\limits_{\scriptstyle  \mathbf{W}_{c,K+1}, f, t
				\atop  \scriptstyle } \ \ \
			& -t \\
			{\rm{s.t.}}\ \ \ \ \ \ \
			& \mathbf{W}_{K+1}, \mathbf{Z}(\mathbf{R}_s)\succeq \mathbf{0}, \\
			& \mathrm{rank}(\mathbf{W}_{c,K+1}) = 1, \\
			&\eqref{p_CRBmin_f}, \eqref{p_CRBmin_C},\eqref{p_CRBmin_P}, \eqref{p_SDR_power}.
		\end{align}
	\end{subequations}
	The second optimization problem w.r.t. other variables is given by
	\begin{subequations} \label{P_AO2}
		\begin{align}
			\!\!\!\!\!\! \mathop  {\min  }\limits_{\scriptstyle \{\mathbf{W}_{c,k}\}_{k=1}^K, t
				\atop   \scriptstyle } \ \ \
			& -t \\
			{\rm{s.t.}}\ \ \ \ \ \ \
			& \mathbf{W}_{c,k}, \mathbf{Z}(\mathbf{R}_s)\succeq \mathbf{0}, \forall 1\leq k \leq K, \\
			& \mathrm{rank}(\mathbf{W}_{c,k}) = 1, \forall 1\leq k \leq K, \\
			& \gamma_p \geq 2^{(R_{p,\min} - f/\rho)/B} - 1, \label{}\\
			& \eqref{p_CRBmin_f}, \eqref{p_CRBmin_C},\eqref{p_SDR_power}.
		\end{align}
	\end{subequations}
	By iteratively solve the two problems \eqref{P_AO1} and \eqref{P_AO2}, we can obtain the local optimal solution for problem \eqref{P1}.
	Note that both the two problems \eqref{P_AO1} and \eqref{P_AO2} are convex by dropping the rank-one constraints. However, additional steps may be required to extract a suboptimal solution. Fortunately, the following theorem ensures that we can always obtain rank-one optimal solutions.
	
\begin{theorem} \label{theorem_point_2}
	Generally, the optimal solutions of both the two problems \eqref{P_AO1} and \eqref{P_AO2} by dropping the rank-one constraints are rank-one.
\end{theorem}
	
{\it{Proof:}} See Appendix B. $\hfill\blacksquare$

\subsection{SDR-Based SCA Approach}
	
The AO method optimizes a subset of variables while keeping the others unchanged which is easily trapped into local optima. Hence, we propose another algorithm based on the SCA technique which jointly optimizes the variables in each iteration. By introducing the slack variable $\gamma_p$, the problem \eqref{P1} can be converted to
\begin{subequations} \label{P1_SCA}
	\begin{align}
		\!\!\!\!\!\!\mathop  {\min  }\limits_{\scriptstyle \{\mathbf{W}_{c,k}\},
				\atop  f, \gamma_p, t \scriptstyle }
		\hspace{0pt} \ \  & -t \\
		{\rm{s.t.}}\ \ \ \;
		&\gamma_p \geq 0, \label{p_c_gammap}\\
		&\frac{ \mathrm{tr}(\mathbf{Q}_{K+1} \mathbf{W}_{c,K+1} )  }{\sum_{i=1}^{K} \mathrm{tr}(\mathbf{Q}_{K+1} \mathbf{W}_{c,i} )  + \sigma^2_c }  \geq \gamma_p , \label{p_C_SCA}\\ 
		&\eqref{p_CRBmin_f}, \eqref{p_CRBmin_C},\eqref{p_SDR_postive}, \eqref{p_SDR_rank_1}, \eqref{p_SDR_power}.
	\end{align}
\end{subequations}
Note that the non-convex constraint \eqref{p_C_SCA} can be further transformed to
\begin{align}  \label{p_C_SCA1}
	\frac{\mathrm{tr}(\mathbf{Q}_{K+1} \mathbf{W}_{c,K+1} )}{\gamma_p} \geq \sum_{i=1}^{K} \mathrm{tr}(\mathbf{Q}_{K+1} \mathbf{W}_{c,i} ) + \sigma^2_c.
\end{align}
We first convert the LHS of \eqref{p_C_SCA1} to the difference of two convex functions, then apply the first-order Taylor expansion to obtain its lower bound. The corresponding process is shown as follows.
\begin{align}
	&2\mathrm{tr}(\mathbf{Q}_{K+1} \mathbf{W}_{c,K+1} ) \frac{1}{ \gamma_p} \nonumber \\
	=& \Big(\mathrm{tr}(\mathbf{Q}_{K+1} \mathbf{W}_{c,K+1} ) + \gamma_p^{-1} \Big)^2 - \Big(\mathrm{tr}^2(\mathbf{Q}_{K+1} \mathbf{W}_{c,K+1} ) + \gamma_p^{-2} \Big) \nonumber \\
	\geq & \Big(\mathrm{tr}(\mathbf{Q}_{K+1} \mathbf{W}_{c,K+1}^{(i)} ) + \frac{1}{ \gamma_p^{(i)}} \Big)^2 - \big(\mathrm{tr}^2(\mathbf{Q}_{K+1} \mathbf{W}_{c,K+1} ) + \gamma_p^{-2} \big) \nonumber \\
	+& 2\Big(\mathrm{tr}(\mathbf{Q}_{K+1} \mathbf{W}_{c,K+1}^{(i)} ) + \frac{1}{ \gamma_p^{(i)}} \Big)\mathrm{tr}(\mathbf{Q}_{K+1} (\mathbf{W}_{c,K+1} -\mathbf{W}_{c,K+1}^{(i)}) ) \nonumber \\
	-& 2(\gamma_p^{(i)})^{-2}(\gamma_p - \gamma_p^{(i)})\Big(\mathrm{tr}(\mathbf{Q}_{K+1} \mathbf{W}_{c,K+1}^{(i)} ) + \frac{1}{ \gamma_p^{(i)}} \Big) \nonumber \\
		\triangleq & 2\hat{f} (\gamma_p, \mathbf{W}_{c,K+1}),
\end{align}
where $\mathbf{W}_{c,K+1}^{(i)}$ and $\gamma_p^{(i)}$ denote the corresponding variables in the $i$-th iteration. Hence, the convex problem can be constructed at each iteration by substituting the surrogate $\hat{f} (\gamma_p, \mathbf{W}_{c, K+1})$ to \eqref{p_C_SCA1}. 
The following theorem ensures that we can always obtain rank-one optimal solutions in each iteration.
	
\begin{theorem} \label{theorem_point_3} 
	If $\mathbf{H} \hat{\mathbf{A}} $ is of full column rank, where $\mathbf{H} = [\mathbf{h}_1,\ldots,\mathbf{h}_{K}, \mathbf{h}_{K+1}]^H \in  \mathbb{C}^{K+1 \times M_t}$, $\hat{\mathbf{A}} = [\mathbf{a}, \dot{\mathbf{a}}]$, we can always obtain the optimal solution $\mathbf{W}_{c,k}, \forall 1 \leq k \leq K+1$ that satisfies the rank-one constraint. Besides, $\mathbf{H} \hat{\mathbf{A}} $ is of full column rank always holds for $K\geq 1$.
\end{theorem}

{\it{Proof:}} See Appendix C. $\hfill\blacksquare$

Note that the iteratively obtained objective function value is non-increasing over iterations by utilizing SDR-based AO and SDR-based SCA algorithms to tackle the original problem. Besides, they are theoretically guaranteed to converge after a certain number of iterations due to that the optimal objective value is lower bound due to the communication and computation requirements as well as the limited power budget at the BS. The worst computational complexity of the SDR-based AO algorithm is given by $\big((M_t+2)^{4.5} + (KM_t+1)^{4.5}\big)I_{AO}$, and the worst computational complexity of SDR-based SCA algorithm is given by $\big((K+1)M_t+3\big)^{4.5}I_{SCA}$ \cite{ZhiquanLuo_SDR}, where $I_{AO}$ and $I_{SCA}$ denote the corresponding number of iterations that needed to converge to a stationary point. One can observe that the SDR-based SCA algorithm is more complicated than the SDR-based AO algorithm generally.
	
\section{Algorithms For Extended Target Case}

Refer to \eqref{CRB2} and define $\widehat{\mathbf{W}}_r = \mathbf{W}_r \mathbf{W}_r^H $, the original CRB minimization problem for the extended target scenario is given by
\begin{subequations} \label{P_extend}
	\begin{align}
		\mathop  {\min  }\limits_{\scriptstyle \{\mathbf{w}_{c,k}\}, \widehat{\mathbf{W}}_r, f 
		\atop \scriptstyle } \ \ \
		&  \mathrm{tr}\bigg(   \Big( \sum_{k=1}^{K+1} \mathbf{W}_{c,k}   + \widehat{\mathbf{W}}_r \Big) ^{-1} \bigg) \\
			{\rm{s.t.}}\ \ \ \ \ \ \
		&  \mathbf{W}_{c,k}, \widehat{\mathbf{W}}_r \succeq \mathbf{0}, \forall 1\leq k \leq K+1, \label{} \\
		& \mathrm{rank}(\mathbf{W}_{c,k}) = 1, \forall 1\leq k \leq K+1, \label{} \\
		& \sum_{k=1}^{K+1} \mathrm{tr}( \mathbf{W}_{c,k}) + \mathrm{tr}( \widehat{\mathbf{W}}_r)  + \kappa f^3 \leq P_{0}, \label{p_ex_SDR_power} \\
		& \eqref{p_CRBmin_f}, \eqref{p_CRBmin_C},\eqref{p_CRBmin_P}.
	\end{align}
\end{subequations}
Note that problem \eqref{P_extend} remains intractable due to rank-one constraint and the coupled variables in constraint \eqref{p_CRBmin_P}. 
We propose both SDR-based AO and SDR-based SCA algorithms similar to the point target case.

\subsection{No Communication User Case}
\label{}
First, we consider the special case that there is no communication requirement, the corresponding problem is given by
\begin{subequations} \label{P_no_comm_e}
	\begin{align}
		\mathop  {\min  }\limits_{\scriptstyle \mathbf{w}_{c,K+1}, \widehat{\mathbf{W}}_r, f
			\atop \scriptstyle } \ \ \
		&  \mathrm{tr}\bigg(   \Big(  \mathbf{W}_{c,K+1}   + \widehat{\mathbf{W}}_r \Big) ^{-1} \bigg) \\
		{\rm{s.t.}}\ \ \ \ \ \ \
		&  \mathbf{W}_{c,K+1}, \widehat{\mathbf{W}}_r \succeq \mathbf{0}, \label{P_no_comm_e1} \\
		& \mathrm{rank}(\mathbf{W}_{K+1}) = 1, \label{P_no_comm_e2} \\
		& \mathrm{tr}( \mathbf{W}_{c,K+1}) + \mathrm{tr}( \widehat{\mathbf{W}}_r)  + \kappa f^3 \leq P_{0}, \label{P_no_comm_e3} \\
		&\hspace{-60pt} f/\rho + B\log_2\Big(1+ \frac{ \mathrm{tr}(\mathbf{Q}_{K+1} \mathbf{W}_{K+1}) }{ \mathrm{tr}(\mathbf{Q}_{K+1} \widehat{\mathbf{W}}_r) + \sigma^2_c }\Big) \geq R_{p,\min}, \label{}\\
		& \eqref{p_CRBmin_f}.
	\end{align}
\end{subequations}

By introducing the slack variable $\gamma_p$, the problem \eqref{P_no_comm_e} can be converted to
\begin{subequations} \label{P_no_comm_e4}
	\begin{align}
		\mathop  {\min  }\limits_{\scriptstyle \mathbf{w}_{c,K+1}, \widehat{\mathbf{W}}_r,
			\atop \scriptstyle f, \gamma_p }
		& \mathrm{tr}\bigg(   \Big( \mathbf{W}_{c,K+1}   + \widehat{\mathbf{W}}_r \Big) ^{-1} \bigg) \\
		{\rm{s.t.}}\ \ \ \;
		& f/\rho + \log_2(1+\gamma_p) \geq R_{p,\min}, \\
		& \frac{ \mathrm{tr}(\mathbf{Q}_{K+1} \mathbf{W}_{c,K+1} )  }{ \mathrm{tr}(\mathbf{Q}_{K+1} \widehat{\mathbf{W}}_r) + \sigma^2_c }  \geq \gamma_p , \label{P2_SCA_c_gammap}\\ 
		& \eqref{p_CRBmin_f}, \eqref{P_no_comm_e1}, \eqref{P_no_comm_e2}, \eqref{P_no_comm_e3}.
	\end{align}
\end{subequations}

\begin{theorem}
	The optimal solutions of the problem \eqref{P_no_comm_e4} by dropping the rank-one constraints are rank-one. 
\end{theorem}

Note that problem \eqref{P_no_comm_e4} is similar to problem \eqref{P_no_comm1}, the corresponding proofs are shown in Appendix A of this paper.

	\subsection{SDR-based AO Approach}
	
	The optimization problem w.r.t.  $\mathbf{w}_{c,K+1},  f$ is given by
	\begin{subequations} \label{P_extend_AO1}
		\begin{align}
			\mathop  {\min  }\limits_{\scriptstyle \mathbf{W}_{c,K+1}, f
				\atop \scriptstyle } \ \ \
			&  \mathrm{tr}\bigg(   \Big( \sum_{k=1}^{K+1} \mathbf{W}_{c,k}   + \widehat{\mathbf{W}}_r \Big) ^{-1} \bigg) \\
			{\rm{s.t.}}\ \ \ \;
			&  \mathbf{W}_{K+1} \succeq \mathbf{0}, \\
			&  \mathrm{rank}(\mathbf{W}_{c,K+1}) = 1, \\
			&  \eqref{p_CRBmin_f}, \eqref{p_CRBmin_C},\eqref{p_CRBmin_P}, \eqref{p_ex_SDR_power}.
		\end{align}
	\end{subequations}
	The optimization problem w.r.t. other variables is given by
	\begin{subequations} \label{P_extend_AO2}
		\begin{align}
			\mathop  {\min  }\limits_{\scriptstyle \{\mathbf{W}_{c,k}\}, \mathbf{W}_{r}
				\atop \scriptstyle } \ \ \
			&  \mathrm{tr}\bigg(   \Big( \sum_{k=1}^{K+1} \mathbf{W}_{c,k}   + \widehat{\mathbf{W}}_r \Big) ^{-1} \bigg) \\
			{\rm{s.t.}}\ \ \ \ \ \ 
			& \mathbf{W}_{c,k} \succeq \mathbf{0}, \forall 1\leq k \leq K, \\
			& \mathrm{rank}(\mathbf{W}_{c,k}) = 1, \forall 1\leq k \leq K, \\
			& \eqref{p_CRBmin_f}, \eqref{p_CRBmin_C},\eqref{p_CRBmin_P}, \eqref{p_ex_SDR_power}.
		\end{align}
	\end{subequations}
	
	Note that the two sub-problems are convex by dropping the non-convex rank-one constraints. It is observed that problem \eqref{P_extend_AO1} is similar to problem \eqref{P_AO1}, and the corresponding rank-one property of the solution can be demonstrated as the process in Appendix A. However, the rank-one property of the variables in \eqref{P_extend_AO2} is hard to obtain, so we introduce new variable $\mathbf{R}_s$ and substitute $\widehat{\mathbf{W}}_r$ as 
	\begin{align} \label{newRs}
		\widehat{\mathbf{W}}_r = \mathbf{R}_s - \sum_{k=1}^{K+1} \mathbf{W}_{c,k}.
	\end{align} 
	The corresponding problem is reformulated as 
	\begin{subequations} \label{P_extend_AO3}
		\begin{align}
			\mathop  {\min  }\limits_{\scriptstyle \{\mathbf{W}_{c,k}\}, \mathbf{R}_{s}
				\atop \scriptstyle } \ \ \
			&  \mathrm{tr}\big( \mathbf{R}_{s}^{-1} \big) \\
			{\rm{s.t.}}\ \ \ \ \ \
			& \mathbf{W}_{c,k}, \mathbf{R}_{s} \succeq \mathbf{0}, \forall 1\leq k \leq K, \\
			& \mathrm{rank}(\mathbf{W}_{c,k}) = 1, \forall 1\leq k \leq K, \\
			& \mathbf{R}_{s} \succeq \sum_{k=1}^{K+1} \mathbf{W}_{c,k}, \\
			& \mathrm{tr}(\mathbf{R}_{s}) \leq P_0, \\
			& \eqref{p_CRBmin_f}, \eqref{p_CRBmin_C},\eqref{p_CRBmin_P},
		\end{align}
	\end{subequations}
	where the constraint \eqref{p_CRBmin_C} is transformed to
	\begin{align} \label{p_CRBmin_CRs}
		(1+\gamma_k^{-1} )\mathrm{tr}(\mathbf{Q}_{k} \mathbf{W}_{c,k} ) \geq \mathrm{tr}(\mathbf{Q}_{k} \mathbf{R}_{s}) + \sigma^2_c, 1 \! \leq \! k \! \leq \! K,
	\end{align}
	and $\gamma_p$ in \eqref{p_CRBmin_P} is given by
	\begin{align}
		\gamma_p = \frac{\mathrm{tr}(\mathbf{Q}_{K+1} \mathbf{W}_{c,K+1}) }{\mathrm{tr}(\mathbf{Q}_{K+1} \mathbf{R}_{s}) -  \mathrm{tr}(\mathbf{Q}_{K+1} \mathbf{W}_{c,K+1} )  + \sigma^2_c }.
	\end{align}
	
	Note that problem \eqref{P_extend_AO3} is convex by dropping the rank-one constraints and can be readily tackled by the existing solvers. However, the sub-optimal solution obtained via direct eigenvalue decomposition or Gaussian randomization will induce additional performance loss. Hence, we resort to another approach that can construct optimal rank-one solutions for the original problem. With any given optimal solutions $\{\mathbf{W}_{c,k}^*\}_{k=1}^{K}$, the rank-one optimal solution with satisfying all the corresponding constraints can be constructed as
		\begin{align}
			\hat{\mathbf{w}}_{k} = (\mathbf{h}_{k}^H \mathbf{W}_{k}^* \mathbf{h}_{k})	\mathbf{W}_{k}^*\mathbf{h}_{k}, 1\leq k \leq K.
		\end{align}
It is observed that the constructed rank-one solution meets all the corresponding constraints without degrading the objective function value. With the obtained transmit covariance $\mathbf{R}_{s}$ and the constructed optimal rank-one transmit beamformer $\{\mathbf{W}_{c,k} \}_{k=1}^K$ for communication users, the dedicated transmit beamformer for radar sensing $\mathbf{W}_{r}$ can be recovered via the Cholesky decomposition.
	
\subsection{SDR-Based SCA Approach}
	
In this section, we propose another algorithm based on the SCA technique. By introducing the slack variable $\gamma_p$, the problem \eqref{P_extend} can be converted to
\begin{subequations} \label{P2_SCA}
\begin{align}
	\mathop  {\min  }\limits_{\scriptstyle \{\mathbf{W}_{c,k}\}, \mathbf{W}_{r}, 
				\atop \scriptstyle f, \gamma_p }
	& \mathrm{tr}\bigg(   \Big( \sum_{k=1}^{K+1} \mathbf{W}_{c,k}   + \widehat{\mathbf{W}}_r \Big) ^{-1} \bigg) \\
			& \hspace{-55pt}{\rm{s.t.}} \ \frac{ \mathrm{tr}(\mathbf{Q}_{K+1} \mathbf{W}_{c,K+1} )  }{\sum_{i=1}^{K} \mathrm{tr}(\mathbf{Q}_{K+1} \mathbf{W}_{c,i} )  + \mathrm{tr}(\mathbf{Q}_{K+1} \widehat{\mathbf{W}}_r) + \sigma^2_c }  \geq \gamma_p , \label{P2_SCA_c_gammap}\\ 
	& \eqref{p_CRBmin_f}, \eqref{p_CRBmin_C}, \eqref{p_CRBmin_P},\eqref{p_SDR_postive}, \eqref{p_SDR_rank_1}, \eqref{p_ex_SDR_power},
\end{align}
\end{subequations}
	
Similar to the reformulation towards problem \eqref{P_extend_AO3}, problem \eqref{P2_SCA} is further transformed to 
	\begin{subequations} \label{P2_SCA1}
		\begin{align}
			\hspace{-20pt} \mathop  {\min  }\limits_{\scriptstyle \{\mathbf{W}_{c,k}\}, \mathbf{R}_{s}, 
				\atop \scriptstyle f, \gamma_p }
			&  \mathrm{tr}\big( \mathbf{R}_{s}^{-1} \big) \\
			{\rm{s.t.}}\ \ \ \;
			& \mathbf{W}_{c,k}, \mathbf{R}_{s} \succeq \mathbf{0}, \forall 1\leq k \leq K+1, \\
			& \mathrm{rank}(\mathbf{W}_{c,k}) = 1, \forall 1\leq k \leq K+1, \\
			& \mathbf{R}_{s} \succeq \sum_{k=1}^{K+1} \mathbf{W}_{c,k}, \\
			& \mathrm{tr}(\mathbf{R}_{s}) \leq P_0, \\
			&\hspace{-20pt} \frac{\mathrm{tr}(\mathbf{Q}_{K+1} \mathbf{W}_{c,K+1}) }{\mathrm{tr}(\mathbf{Q}_{K+1} \mathbf{R}_{s}) -  \mathrm{tr}(\mathbf{Q}_{K+1} \mathbf{W}_{c,K+1} )  + \sigma^2_c } \geq \gamma_p , \label{P2_SCA_c_gammap}\\ 
			& \eqref{p_CRBmin_f}, 
			\eqref{p_CRBmin_P},\eqref{p_SDR_rank_1}, \eqref{p_CRBmin_CRs},
		\end{align}
	\end{subequations}
	in each iteration, the constraint \eqref{P2_SCA_c_gammap} can be replaced by 
	\begin{align}
		\hat{f} (\gamma_p, \mathbf{W}_{c,K+1}) \geq & \mathrm{tr}(\mathbf{Q}_{K+1} \mathbf{R}_{s}) \nonumber \\ 
		&- \mathrm{tr}(\mathbf{Q}_{K+1} \mathbf{W}_{c,K+1} ) + \sigma^2_c. 
	\end{align}
Hence, problem \eqref{P2_SCA1} can be tackled by solving a series of convex optimization problems iteratively and finally converges after sufficient iterations. Similarly as the analysis for problem \eqref{P_extend_AO3}, with any given optimal solutions $\{\mathbf{W}_{c,k}^*\}_{k=1}^K+1$, we can construct the rank-one optimal solution for $1 \leq k \leq K+1$ as $\hat{\mathbf{w}}_{k} = (\mathbf{h}_{k}^H \mathbf{W}_{k}^* \mathbf{h}_{k})	\mathbf{W}_{k}^*\mathbf{h}_{k}$, and further construct the dedicated transmit beamformer for radar sensing $\mathbf{W}_{r}$ with the obtained optimal variables $\{\mathbf{W}_{c,k}^*\}_{k=1}^{K+1}$ and $\mathbf{R}_s$.
	
Similar to the analysis of the point target scenario, the proposed two algorithms are theoretically guaranteed to converge after a certain number of iterations. The corresponding worst computational complexity of the two algorithms are given by $\big((M_t+1)^{4.5} + ((K+1)M_t)^{4.5}\big)I_{AO}$, and $\big((K+2)M_t+2\big)^{4.5}I_{SCA}$, respectively. 


\section{Simulation Results}
	
In this section, we evaluate the performance of the proposed SDR-based AO and SDR-based SCA algorithms for the ISCC system via extensive numerical results. The BS and server are located at $(0,0,20)$ meter (m), and $(20,20,5)$ m, respectively. The placement of communication UEs is random within a radius of $10$ m that is centered at $(30,30,0)$ m, and the distance between the BS and target is set as $50$ m. The multi-function BS is equipped with $M_t=16$ transmit antennas and $M_r=20$ receive antennas. 
The distance-dependent path loss for communication links is modeled as $PL(d) = \rho_0 (d/d_0)^{-\alpha }$, where $\rho_0$ corresponds to the path loss at the reference distance $d_0=1$ m and is given as $\rho_0 = -30$ dB \cite{Guangji_active}. 
Besides, $d$ denotes the link distance and $\alpha$ denotes the path loss exponent. The path-loss exponents $\alpha$ of communication links are set to $3$, and the Rician factors are set as $5$ dB.
Besides, the noise power at the BS, server, and the UEs are set as $\sigma^2 = -110$ dBm. Unless otherwise stated, other parameters are set as: $B = 1$ MHz, $\rho= 1000$ cycles/bit, $\kappa = 10^{-28}$ \cite{GuangjiChen_DIBF_MEC}. For the point target scenario, we assume that the target angle is $\theta=0^{\circ}$ and the radar cross section is $3 \ \mathrm{m}^2$.
 For comparison, we consider the following schemes: 1) \textbf{Approx. BP}: the beampattern approximation scheme proposed in \cite{XiangLiu_TSP20_MIMORadar}; 2) \textbf{SDR-based AO bound}: relaxing the rank-one constraint and without recovering the original transmit precoder in the SDR-based AO scheme, which serves as a performance upper bound; 3) \textbf{SDR-based SCA bound}: similar as the process in 2) for the SDR-based SCA scheme; 4) \textbf{offloading-only}: the BS purely offload the task to the server without local computing.
	
\subsection{Point Target Case}

\begin{figure}[t]  
	\centering
	\includegraphics[width=0.8\columnwidth]{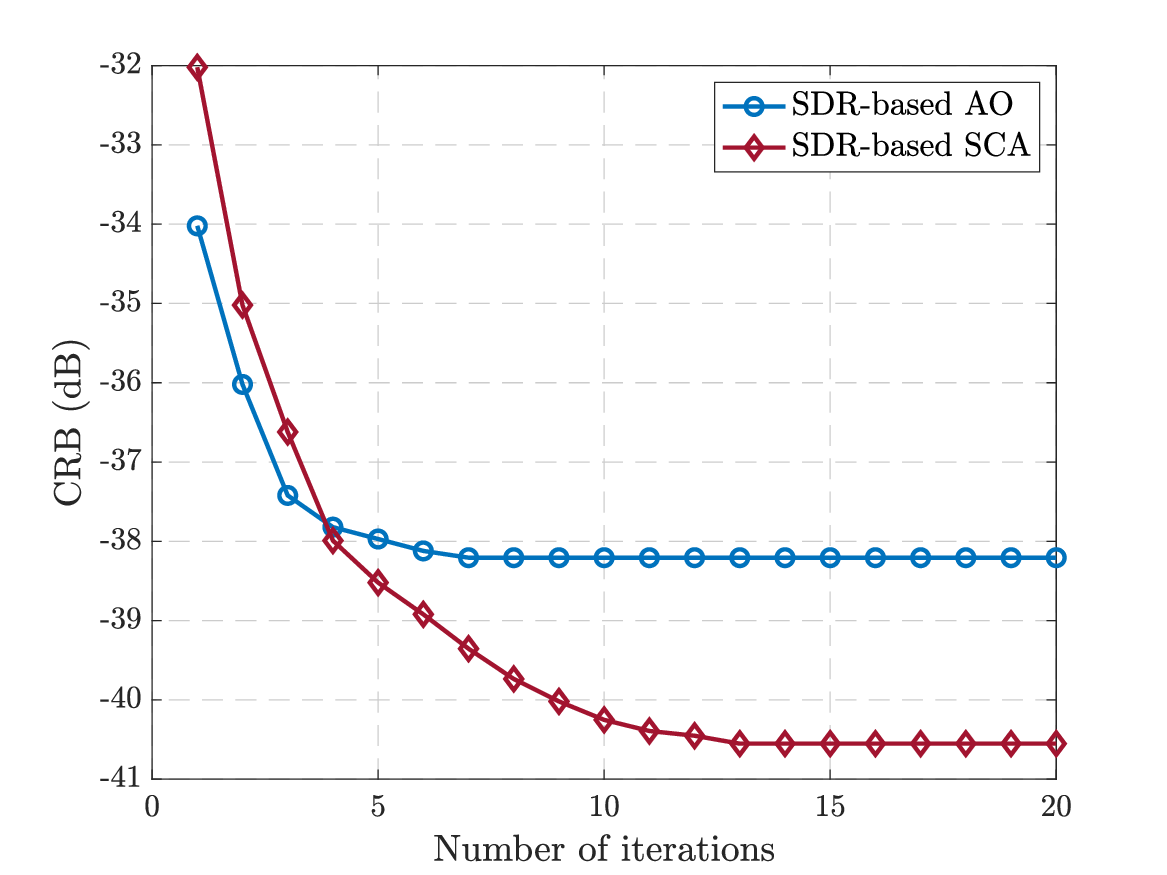} 
	\caption{The convergence performance.}
	\label{convergence}	
\end{figure}

\begin{figure}[t]  
	\centering
	\includegraphics[width=0.8\columnwidth]{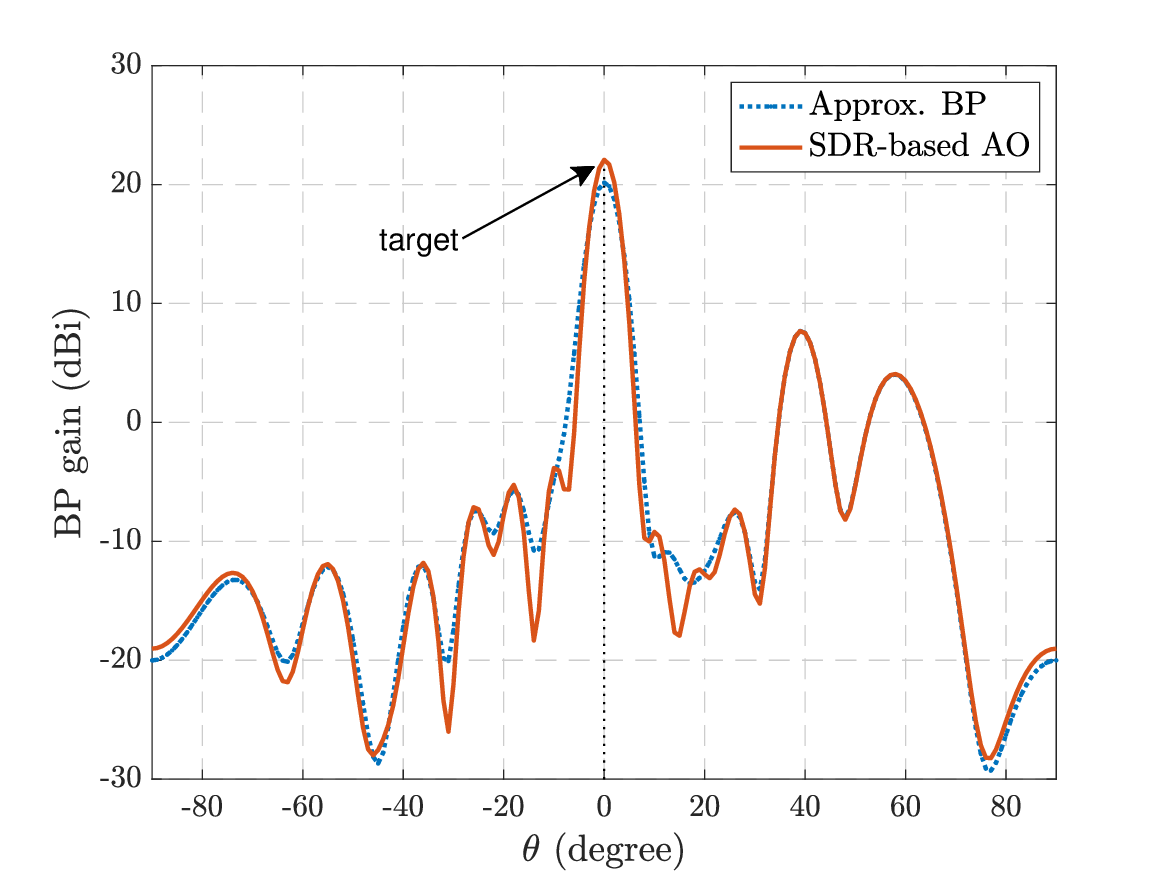} 
	\caption{The beampattern performance.}
	\label{fig_BP}	
\end{figure}
At the beginning, we analyze and compare the convergence behavior of the proposed algorithms under the point target case. The transmit power budget at the BS and the SINR requirement at the communication users are set as $30$ dBm, and $20$ dB, respectively.
Besides, the computation requirement is set as $1$ Mbits. Fig. \ref{convergence} shows that the achieved CRBs of both algorithms decrease rapidly with the number of iterations. One can observe that the SDR-based SCA algorithm can achieve lower CRB but requires more iterations than the SDR-based AO algorithm. This is due to that the SCA technique can jointly optimize the variables iteratively.
Then, we compared the obtained transmit beamforming via the SDR-based AO method to the Approx. BP scheme as an example. The reference beampattern with $3$ dB beamwidth of $10^{\circ}$ for radar sensing can obtained via the standard approach in \cite{JianLi_SPMag07}. The number of communication users is $K=3$, while the other parameters remain the same.
It can be seen that the beampatterns precisely align the main lobes toward the target angle $\theta=0^{\circ}$, and the leakage energy in sidelobe regions is intelligently utilized to meet the SINR constraints for users and the task offloading requirement on the BS. 
We can observe that the proposed algorithms can achieve the maximum beampattern gain while meeting the communication and computation requirements.

\begin{figure}[t]  
	\centering
	\includegraphics[width=0.8\columnwidth]{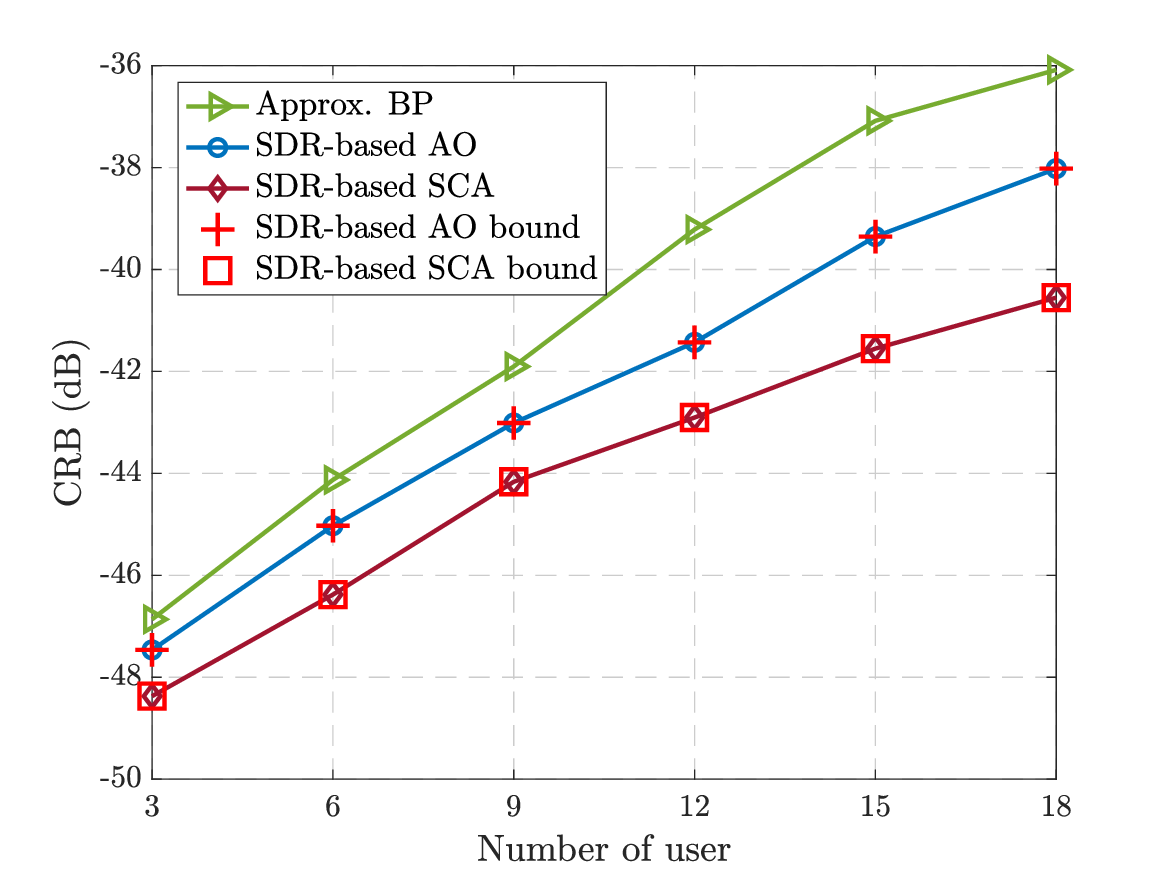} 
	\caption{CRB versus the number of users.}
	\label{fig_K}
\end{figure}
	
Next, we compare the achieved CRB  versus the number of communication users under the point target case. 
Notice that with the growing number of communication users, the achieved CRB also tends to increase. This is primarily due to that the transmit power and spatial diversity need to be enlarged to support more communication users. 
Besides, it is observed that the gap between different algorithms is becoming large, especially the beampattern approximation scheme. This demonstrates that the proposed algorithms can obtain higher target estimation performance, as compared to conventional beampattern approximation approaches. 
Fortunately, the variation of the CRB can be kept within $2$ dB when $K=3$, which demonstrates that the transmit signal beam for communication users can be utilized to precisely sense the target with negligible performance loss when the number of communication users is small. When the number of users becomes large, the ISCC system can still maintain a satisfying service level. Besides, it is observed that the obtained solution via eigenvalue decomposition can achieve the same performance compared to the performance bound without recovering the original transmit beamformer. 
This verifies the proved rank-one property of the obtained solutions in the proposed theorems and remarks.
	
\begin{figure}[t]  
	\centering
	\includegraphics[width=0.8\columnwidth]{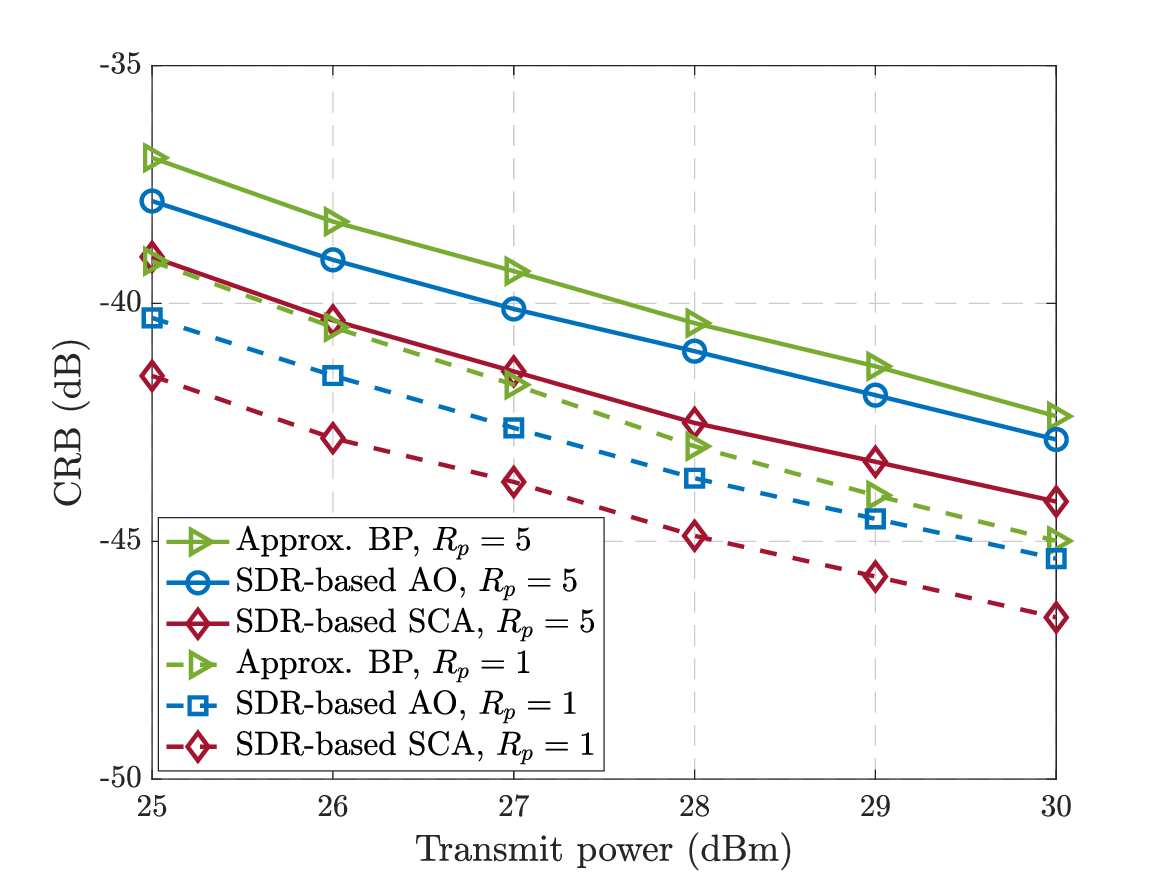} 
	\caption{CRB versus the transmit power.}
	\label{fig_vsP0}
\end{figure}
	
Fig. \ref{fig_vsP0} presents the curve that shows the achieved CRB versus the available transmit power $P_0$ when $R_p=1$ and $R_p=5$, where the corresponding SINR threshold for the communication requirement is set as $25$ dB.
First, it is observed that both the proposed SDR-based AO and SDR-based SCA algorithms can achieve lower CRB compared to the beampattern approximation method in \cite{XiangLiu_TSP20_MIMORadar}, and the SDR-based SCA algorithm is better the SDR-based AO algorithm. This observation further demonstrates the effectiveness of the proposed algorithm.
Second, one can see that the CRB for estimating the point target decreases linearly with the transmit power $P_0$ since they are inversely proportional, more available transmit power can provide more sensing accuracy.
Furthermore, we can observe that the proposed algorithm when $R_p=5$ can achieve $94.77\%$ performance compared to $R_p=1$, but can achieve $5$ times of computation rate.

\begin{figure}[t]  
	\centering
	\includegraphics[width=0.8\columnwidth]{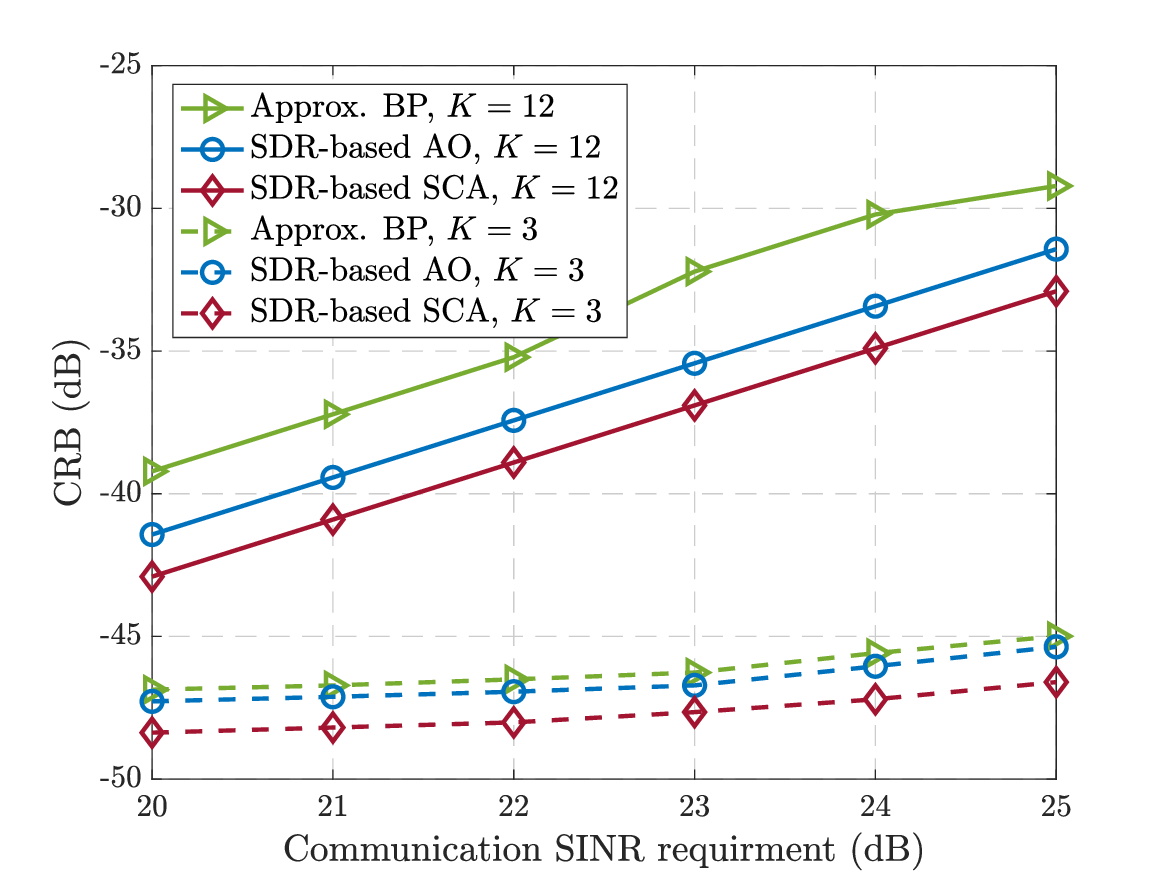} 
	\caption{CRB versus SINR requirement.}
	\label{fig_QoS}
\end{figure}
	
In Fig. \ref{fig_QoS}, we consider the performance tradeoff between radar and communication. 
One can see that the achieved CRB increases progressively with the increase of communication SINR threshold. For a smaller number of users, e.g., $K=3$, the CRB remains at a low level despite that the required communication SINR is growing. When a large number of users need the communication service, e.g., $K=12$, the CRB increases fast, i.e., the sensing accuracy degrades fast since the base station needs to allocate more power to combat the path loss and mitigate the interference among destinations. Besides, it is noticed that the performance gap between the traditional approximation method and the proposed algorithm in this paper becomes large when the number of users increases. This again demonstrates the superiority of the proposed algorithm.
	
\begin{figure}[t]  
	\centering
	\includegraphics[width=0.8\columnwidth]{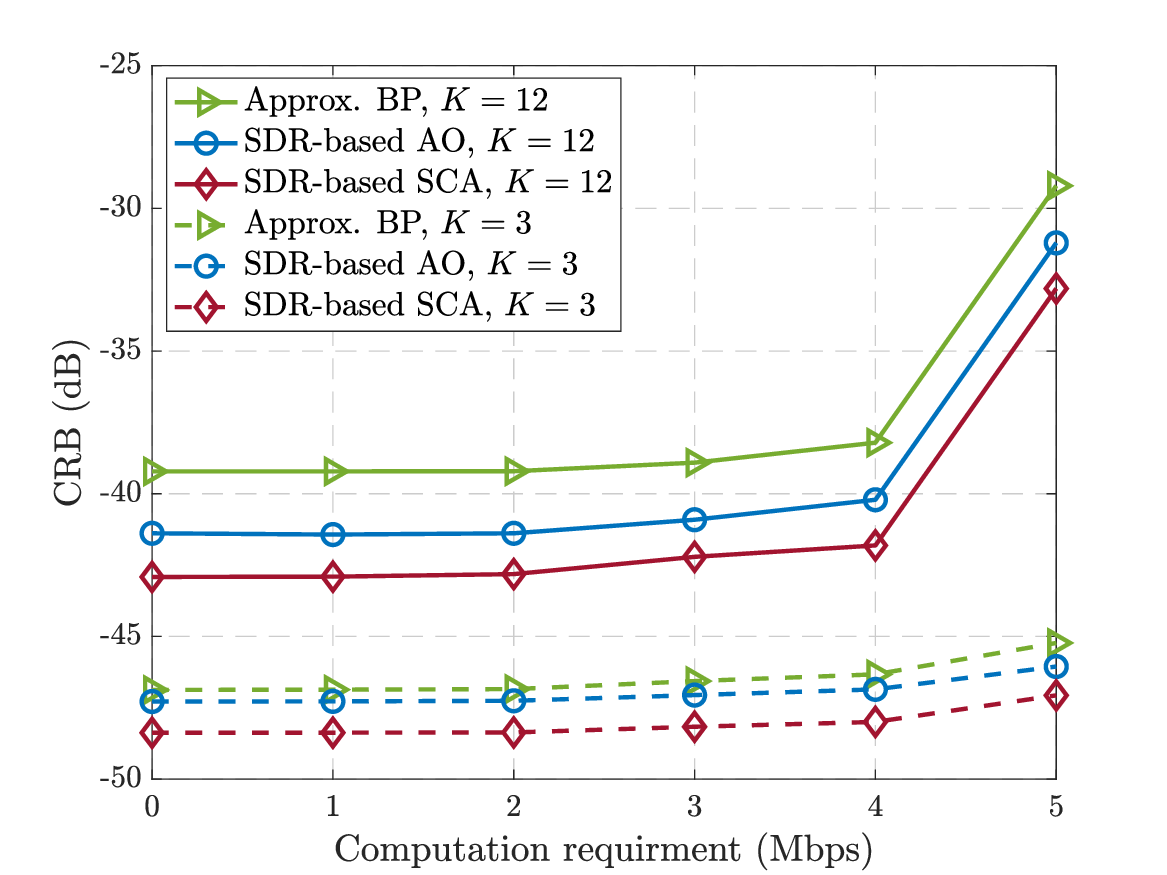} 
	\caption{CRB versus the computation requirement.}
	\label{fig_Rp}
\end{figure}

In Fig. \ref{fig_Rp},  we compare the achieved CRB versus the computation requirement. Fortunately, the variation of the CRB can be kept within 1 dB when $K=3$. However, when the number of users increases to $K=12$, the CRB increases sharply with the computation bits requirement increases. The inherent reason is similar to the increased requirement of communication, more computation requirement needs a faster task offloading rate and larger local computing rate.

\subsection{Extended Target Case}

\begin{figure}[t]  
	\centering
	\includegraphics[width=0.8\columnwidth]{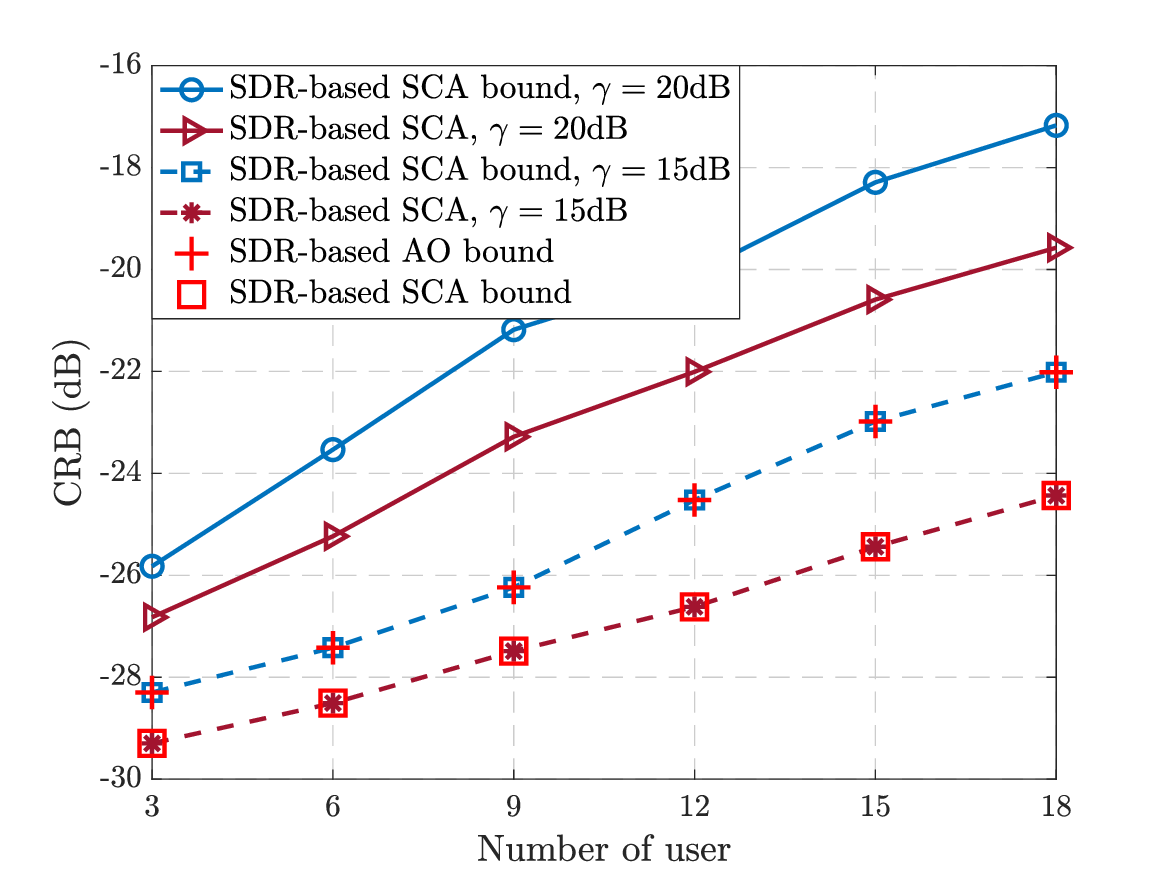} 
	\caption{CRB versus the number of users under the extended target case.}
	\label{fig_Ke}
\end{figure}
	
In Fig. \ref{fig_Ke}, we depict the achieved CRB for estimating the entire TRM $\mathbf{G}$ versus the number of users under the extended target scenario. It is observed that, similarly to the point target case, the achieved CRB is in direct proportion to the number of users. This is because more communication users need more resources, e.g., power and spatial resources. Besides, it is observed that the constructed rank-one solution achieves the same performance compared to the performance bound without recovering the original transmit beamformer. This demonstrates that the proposed algorithms can achieve the best performance with low computation complexity.
	
\begin{figure}[t]  
	\centering
	\includegraphics[width=0.8\columnwidth]{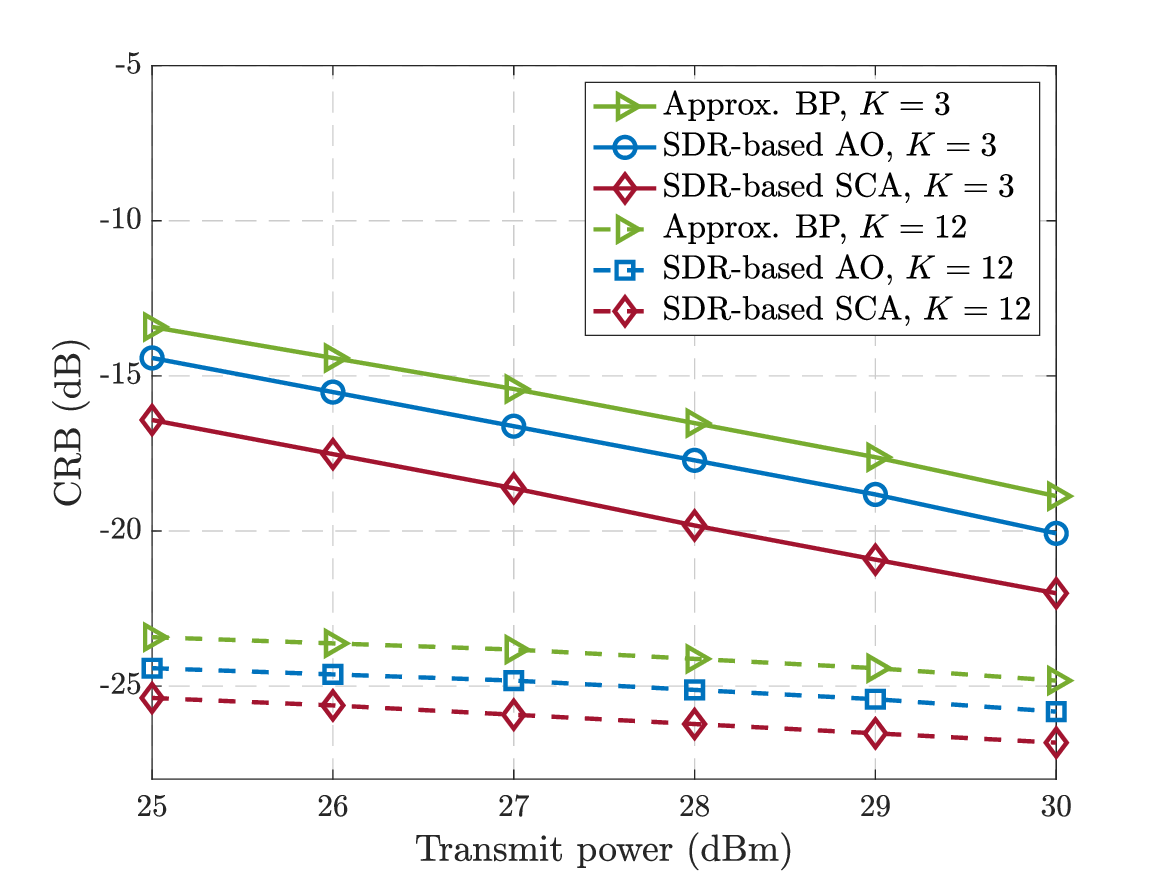} 
	\caption{CRB versus the number of users under the extended target case.}
	\label{fig_P0e}
\end{figure}

Fig. \ref{fig_P0e} shows the achieved CRB versus the available transmit power on the BS when $K=3$ and $K=12$, where the corresponding SINR threshold for the communication requirement is set as $20$ dB.
It is observed that, similar to the point target case, both the proposed SDR-based AO and SDR-based SCA algorithms can achieve lower CRB compared to the beampattern approximation method, and the SDR-based SCA algorithm achieves the best performance. Besides, the achieved CRB gradually decreases with the increase of transmit power $P_0$. Furthermore, we can observe that the CRB can maintain a satisfying performance with fewer communication users even if the avaiable transmit power is limited. 
These observations further demonstrate the effectiveness of the proposed algorithm.

\begin{figure}[t]  
	\centering
	\includegraphics[width=0.8\columnwidth]{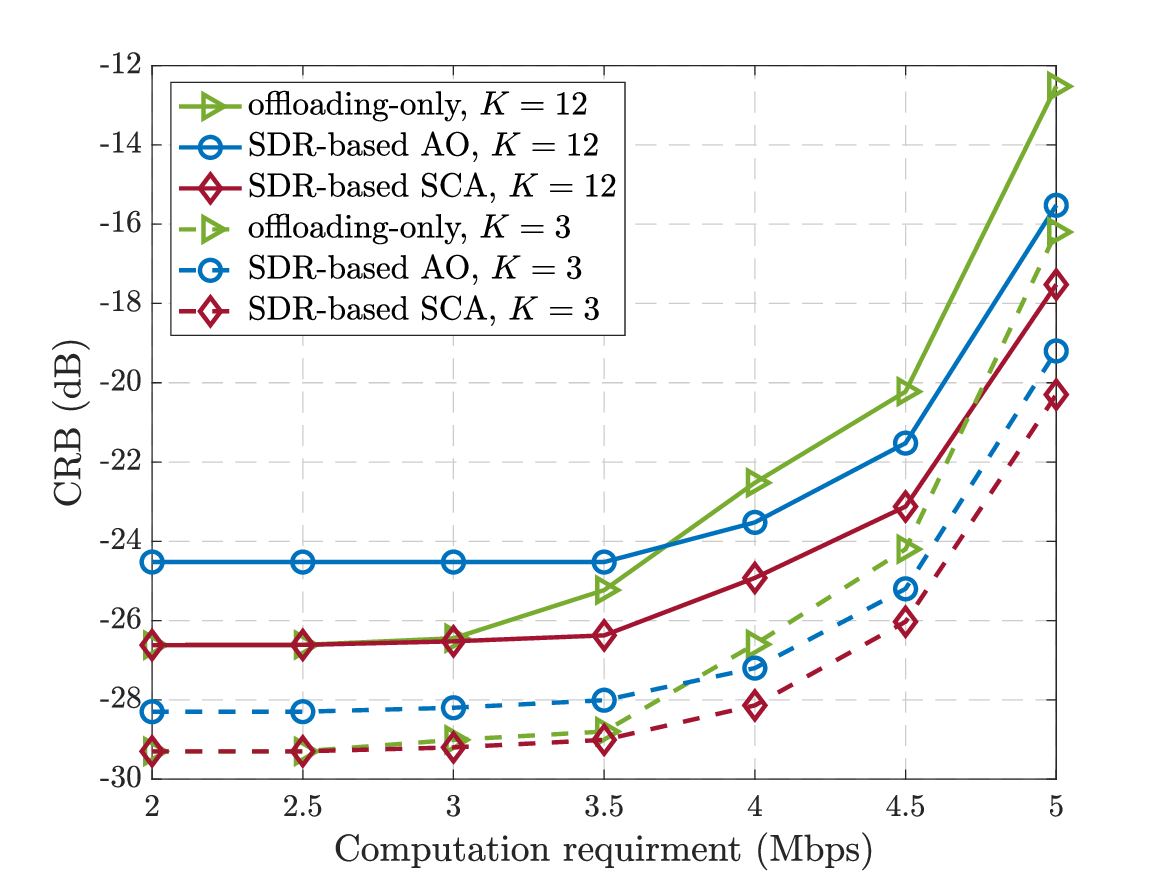} 
	\caption{CRB versus the computation requirement under the extended target case.}
	\label{fig_Rpe}
\end{figure}
	
In Fig. \ref{fig_Rpe}, we compared the performance to the offloading-only strategy under different computation requirements and different numbers of communication users. It can be seen that the pure offloading strategy achieves the best performance as the SDR-based SCA method when computation requirment is low. However, when the computation demand becomes higher, it falls behind quickly.
This is mainly due to that the task offloading signal will cause higher interference to the communication users and the target if the BS still fully offloads the tasks to the edge server.
Note that the pure local computing case cannot satisfy the power and communication constraints when $R_p$ is large, and its performance will be the worst. Hence, we omit the performance comparison in the figure.

\section{Conclusion}
In this paper, we proposed a design framework for the ISCC system which aims to simultaneously provide sensing, communication, and computation services. To tackle the non-convex optimization problem, we proposed both SDR-based AO and SDR-based SCA algorithms to minimize the CRB of radar sensing while guaranteeing the SINR for communication users and the requirement of computation at the BS. Then, we demonstrated that the original problem can be optimally tackled through semidefinite programming with dropped rank-one constraints. Furthermore, numerical results verified this finding, thus providing evidence that the proposed algorithms are efficient and have no performance loss. 
It was observed that both the SDR-based AO and SDR-based SCA algorithms can achieve lower CRB compared to the traditional beampattern approximation approach. This paper demonstrated that the ISCC system can simultaneously provide these three functionalities via precise power allocation and beamforming design.

\section*{Appendix A: \textsc{Proof of Theorem \ref{theorem_point_1} }}
	
The corresponding Lagrange function for problem \eqref{P_no_comm1} is given by
\begin{align} 
	&\mathcal{L}(\{\mathbf{X}\}, \{\mathbf{\Phi}\}) =
		-\mathrm{tr}(\mathbf{A} \mathbf{W}_{K+1}) - \eta_1 f + \eta_2(f-f_{\max}) \nonumber \\ 
	&- \eta_3 \Big(f/\rho + B\log_2\Big(1+ \frac{ \mathrm{tr}(\mathbf{H} \mathbf{W}_{K+1}) }{ \sigma^2_c }\Big) - R_{p,\min}\Big) \nonumber \\
	&+ \eta_4(\mathrm{tr}(\mathbf{W}_{K+1}) + \kappa f^3 - P_0) - \mathrm{tr}\big( {\mathbf{\Phi}_{K+1} \mathbf{W}_{c,K+1}}\big),
\end{align}
where $\{\eta\}$ is the non-negative dual variables for the constraints and $\mathbf{\Phi}_{K+1}$ is the semidenite dual variables for the semidenite constraint. 
According to the KKT condition, we obtain that
\begin{align}
	&\mathbf{\Phi}_{K+1} = \mu \mathbf{I} - \Big( \underbrace{ \mathbf{A} + \frac{\eta_3 B }{\ln 2} \frac{1}{1+\mathrm{tr}(\mathbf{H} \mathbf{W}_{K+1}/\sigma_c^2)} \mathbf{H}\Big)} _{\mathbf{A}'}, \\
	&\mathrm{tr}\big( {\mathbf{\Phi}_{K+1} \mathbf{W}_{c,K+1}}\big) = 0. \label{Appendix_noCom_1}
\end{align}

\begin{lemma}
	The eigenvalues of $\mathbf{Z} = \mathbf{x}\mathbf{x}^H + \mathbf{y}\mathbf{y}^H$ is the same only when 
	\begin{align}
		| \mathbf{x} | =  |\mathbf{y}|,\  \mathbf{x}^H \mathbf{y} = 0.
	\end{align}
\end{lemma}

{\it{Proof:}} We have 
\begin{align}
	\mathbf{Z}^2 = \mathbf{x}\mathbf{x}^H \mathbf{x}\mathbf{x}^H + 2 \mathbf{x}\mathbf{x}^H \mathbf{y}\mathbf{y}^H + \mathbf{y}\mathbf{y}^H\mathbf{y}\mathbf{y}^H,
\end{align}
	hence, 
	\begin{align}
		&\mathrm{tr}(\mathbf{Z}) = |\mathbf{x}|^2 + |\mathbf{y}|^2, \nonumber \\
		&\mathrm{tr}(\mathbf{Z}^2 ) = |\mathbf{x}|^4 + |\mathbf{y}|^4 + 2  (\mathbf{x}^H \mathbf{y})^2.
	\end{align}
	Furthermore, we have the eigenvalues $\lambda_1, \lambda_2$ of $\mathbf{A}$ satisfy	
	\begin{align}
		&\lambda_1 + \lambda_2  = |\mathbf{x}|^2 + |\mathbf{y}|^2, \nonumber \\
		&\lambda_1 \lambda_2 =  |\mathbf{x}|^2 |\mathbf{y}|^2 -  (\mathbf{x}^H \mathbf{y})^2,
	\end{align}
	So $\lambda_1, \lambda_2$  are the roots of $\mathbf{A}^2 -  (|\mathbf{x}|^2 + |\mathbf{y}|^2)\mathbf{A} + |\mathbf{x}|^2 |\mathbf{y}|^2- (\mathbf{x}^H \mathbf{y})^2$, and the discriminant is
	\begin{align}
		\Delta = (|\mathbf{x}|^2 - |\mathbf{y}|^2)^2 + 4 (\mathbf{x}^H \mathbf{y})^2.
	\end{align}
It is observed that $\Delta=0$ only when $|\mathbf{x}| =  |\mathbf{y}|$ and $\mathbf{x}^H \mathbf{y} = 0$. Hence completes the proof. $\hfill\blacksquare$

According to Lemma 2, we can obtain that the two eigenvalues of $\mathbf{A}'$ are generally not the same. From \eqref{Appendix_noCom_1}, we have
\begin{align}
		\mathrm{rank}(\mathbf{W}_{c,k}) \leq \mathrm{dim}(\mathcal{N}(\mathbf{\Phi}_k)) ) = M_t - \mathrm{rank}(\mathbf{\Phi}_k).
\end{align}
Hence, we can observe that $\mathrm{rank}(\mathbf{W}_{c,K+1}) = 1 $ in general. Besides, we can refer to Theorem 3.2 in \cite{Rank1_TSP10}, problem \eqref{P_no_comm1} always has an optimal solution $\mathbf{W}_{K+1}$ satisfying $\mathrm{rank}^2(\mathbf{W}_{K+1}) \leq 2$. Hence, there always exists an optimal solution $\mathbf{W}_{K+1}$ satisfying $\mathrm{rank}(\mathbf{W}_{K+1}) = 1$.
	
	\section*{Appendix B: \textsc{Proof of Theorem \ref{theorem_point_2} }}
		
	First, there always exists an optimal solution $\mathbf{W}_{K+1}$ satisfying $\mathrm{rank}(\mathbf{W}_{K+1}) = 1$ for problem \eqref{P_AO1}. Note that problem \eqref{P_AO1} is similar to problem \eqref{P_no_comm1}, and the corresponding proof is shown in Appendix A of this paper.
	Second, from Appendix C in \cite{FanLiu_TSP22_CRB}, we can obtain the following result: if $\mathbf{H} \hat{\mathbf{A}} $ is of full column rank, where $\mathbf{H} = [\mathbf{h}_1,\ldots,\mathbf{h}_{K}, \mathbf{h}_{K+1}]^H \in  \mathbb{C}^{K+1 \times M_t}$, $\hat{\mathbf{A}} = [\mathbf{a}, \dot{\mathbf{a}}]$, the optimal solution $\mathbf{W}_{c,k}, \forall 1 \leq k \leq K+1$ obtained by dropping the rank-one constraint in problem \eqref{P_AO2} always satisfies $\mathrm{rank}(\mathbf{W}_{c,k}) = 1$. Note that the considered channel is random in which the entries are independently distributed, hence, we have $\mathbf{H} \hat{\mathbf{A}} $ is of full column rank when $K\geq 1$.

\section*{Appendix C: \textsc{Proof of Theorem \ref{theorem_point_3} }}
	
	Denote $\eta_f, \eta_p, \{\eta_k\}, \nu, \mu$ as non-negative dual variables for the linear constraints, and $\mathbf{\Phi}_p, \{\mathbf{\Phi}_k\}$ as semidefinite dual variables for the semidefinite constraints, the corresponding Lagrange function is given by
	\begin{align} \label{Appendix_Lagrange}
		& \mathcal{L}(\{\mathbf{X}\}, \{\mathbf{\Phi}\}) = 
		-t - \eta_f f - \eta_p \gamma_p- \sum_{k=1}^{K+1} \mathrm{tr}(\mathbf{\Phi}_k \mathbf{W}_{c,k})  \nonumber \\ 
		& - \mathrm{tr}\big( {\mathbf{\Phi}_p \mathbf{Z}(\mathbf{R}_s)}\big) - \sum_{k=1}^K \eta_k \bigg( \mathrm{tr}(\mathbf{Q}_k \mathbf{W}_{c,k} ) \nonumber \\ 
		&  -  \gamma_{k} \Big(\sum_{i=1, i\neq k}^{K+1} \mathrm{tr}(\mathbf{Q}_k \mathbf{W}_{c,i} ) + \sigma^2_c \Big) \bigg) \nonumber \\
		& - \nu \bigg( \hat{f} (\gamma_p, \mathbf{W}_{c,K+1}) - \sum_{i=1}^{K} \mathrm{tr}(\mathbf{Q}_{K+1} \mathbf{W}_{c,i} ) -\sigma_c^2 \bigg)  \nonumber \\
		& + \mu \bigg( \sum_{k=1}^{K+1} \|\mathbf{w}_{c,k}\|^2 +  \|\mathbf{W}_{r}\|_F^2 + \kappa f^3 - P_{0} \bigg),
	\end{align}

	%
	%
	
	The corresponding first-order derivatives to the variables are given by
	\begin{align}
		&\frac{\partial \mathcal{L}}{\partial \mathbf{W}_{c,k}} = -\mathbf{\Phi}_k - \mathbf{F} - \eta_k(1 +\gamma_k ) \mathbf{Q}_k  + \sum_{i=1}^{K} \eta_i \gamma_i \mathbf{Q}_i  \nonumber \\ 
		& \qquad \qquad + \nu  \mathbf{Q}_{K+1} + \mu \mathbf{I},\ \forall 1\leq k \leq K, \\
		&\frac{\partial \mathcal{L}}{\partial \mathbf{W}_{c,K+1}} = -\mathbf{\Phi}_{K+1} - \mathbf{F} + \sum_{k=1}^K \eta_k \gamma_k \mathbf{Q}_k - \nu \frac{\partial \hat{f} }{\partial \mathbf{W}_{c,K+1}} + \mu \mathbf{I} \nonumber \\
		&=-\mathbf{\Phi}_{K+1} - \mathbf{F} + \sum_{k=1}^K \eta_k \gamma_k \mathbf{Q}_k - \nu \Big(\mathrm{tr}(\mathbf{Q}_{K+1} \mathbf{W}_{c,K+1}^{(i)} ) \nonumber \\
		& - \mathrm{tr}(\mathbf{Q}_{K+1} \mathbf{W}_{c,K+1} )+ \frac{1}{ \gamma_p^{(i)}} \Big) \mathbf{Q}_{K+1} + \mu \mathbf{I},
	\end{align}
	where $\mathbf{F} = \frac{\partial \mathrm{tr}\big( {\mathbf{\Phi}_p \mathbf{Z}(\mathbf{R}_s)}\big)}{\mathbf{W}_k}, 1\leq k \leq K+1 $ is a $\mathrm{rank}$-2 matrix.
	Thus we have
	\begin{align}
		& \mathbf{\Phi}_k = \mu \mathbf{I} - \mathbf{F}_k',\ \forall 1\leq k \leq K+1, 
	\end{align}
	where
	\begin{align}
		& \mathbf{F}_k' = \mathbf{F} + \eta_k(1 +\gamma_k ) \mathbf{Q}_k  -\sum_{i=1}^{K} \eta_i \gamma_i \mathbf{Q}_i \nonumber \\
		&\qquad - \nu  \mathbf{Q}_{K+1},\ \forall 1\leq k \leq K, \\
		&\mathbf{F}_{K+1}' = \mathbf{F} - \sum_{k=1}^K \eta_k \gamma_k \mathbf{Q}_k + \mathbf{C}_{K+1},
	\end{align}
	where
	\begin{align}
		\mathbf{C}_{K+1} = & \nu \Big(\mathrm{tr}(\mathbf{Q}_{K+1} \mathbf{W}_{c,K+1}^{(i)} ) - \mathrm{tr}(\mathbf{Q}_{K+1} \mathbf{W}_{c,K+1} ) \nonumber \\
		&+\frac{1}{ \gamma_p^{(i)}} \Big) \mathbf{Q}_{K+1}.
	\end{align}
	
	According to \eqref{Appendix_Lagrange}, the complementary conditions are
	\begin{align}
		\mathrm{tr}(\mathbf{\Phi}_k \mathbf{W}_{c,k}) 
		= \mathrm{tr}(\mathbf{\Phi}_c \widehat{\mathbf{W}}_r) 
		= \mathrm{tr}\big( {\mathbf{\Phi}_p \mathbf{Z}(\mathbf{R}_s)}\big) = 0.
	\end{align}
	We can further obtain that $\mathrm{rank}(\mathbf{\Phi}_k)\leq M_t-1$ since $\mathbf{\Phi}_k$ and $\mathbf{W}_{c,k}$ are semidefinite matrices. Besides, we have
	\begin{align}
		\mathrm{rank}(\mathbf{W}_{c,k}) \leq \mathrm{dim}(\mathcal{N}(\mathbf{\Phi}_k)) ) = M_t - \mathrm{rank}(\mathbf{\Phi}_k).
	\end{align}

	The rank-one property of the obtained or further constructed optimal solution for the case when $\nu = 0$ has been demonstrated in \cite{FanLiu_TSP22_CRB}. In the scenario where $\nu > 0$, we divide it into various cases and further analyze the structure of the obtained solution. We first split the communication users into two subsets, and then discuss different cases. Denote $\mathcal{K}_1 \triangleq \{k | \eta_k > 0, \forall k \leq K\} = \{1,\ldots, M\}$, and $\mathcal{K}_2 \triangleq \{k | \eta_k = 0, \forall k \leq K\} = \{M+1,\ldots, K\}$. Let $\hat{\mathbf{H}}=[\mathbf{h}_1, \ldots, \mathbf{h}_M]^H$.
	
	\subsubsection{\textbf{Case 1}} $\mathcal{K}_1 = \emptyset$.
	In that case, $\eta_k=0, 1\leq k \leq K$. We have 
	\begin{align}
		&\mathbf{F}_k' = \mathbf{F} - \nu  \mathbf{Q}_{K+1}, \\
		&\mathbf{F}_{K+1}' = \mathbf{F} + \mathbf{C}_{K+1}.
	\end{align}
	Due to the property of positive semidefinite matrix, it is observed that $\mathbf{F}_k'$ only have one maximum eigenvalue, and $\mathbf{F}_{K+1}'$ may have two maximum eigenvalues if $\mathrm{tr}(\mathbf{Q}_{K+1} \mathbf{W}_{c,K+1}^{(i)} ) -\mathrm{tr}(\mathbf{Q}_{K+1} \mathbf{W}_{c,K+1}) + 1/ \gamma_p^{(i)} \geq 0$. Hence,
	\begin{align}
		& \mathrm{rank}(\mathbf{W}_{c,k}) = 1, 1 \leq k \leq K \\
		&M_t - 2 \leq \mathrm{rank}(\mathbf{\Phi}_{K+1}) \leq M_t - 1.
	\end{align}
	Besides, we have
	\begin{align}
		\mu = \lambda_{\max}(\mathbf{F}_k') = \lambda_{\max}(\mathbf{F}_{K+1}').
	\end{align}
	The equation holds only if $\mathbf{h}_{K+1}^H \mathbf{f}_{\max} = 0 $, where $\mathbf{f}_{\max} $ is the eigenvector of $\mathbf{F} $ corresponding to the largest eigenvalue $\lambda_1$. Thus, we obtain that 
	\begin{align}
		&\mathbf{W}_{c,k} = a_k \mathbf{f}_{\max} \mathbf{f}_{\max}^H, \\
		&\mathbf{W}_{c,K+1} = a_{K+1} \mathbf{f}_{\max} \mathbf{f}_{\max}^H +  b_{K+1} \mathbf{h}_{K+1} \mathbf{h}_{K+1}^H. 
	\end{align}
	We can further construct the rank-one solution as
	\begin{align}
		&\mathbf{W}_{c,1}' = (a_1 +  a_{K+1} )\mathbf{f}_{\max} \mathbf{f}_{\max}^H, \\
		&\mathbf{W}_{c,K+1}' =  b_{K+1} \mathbf{h}_{K+1} \mathbf{h}_{K+1}^H,
	\end{align}
	for the original problem.
	It can be readily verified that the constructed solutions are optimal for the original problem and have rank-one property.


	\subsubsection{\textbf{Case 2}} $| \mathcal{K}_1 | = 1$. In that case,
	\begin{align}
		& \mathbf{F}_1' = \mathbf{F} + \eta_1 \mathbf{Q}_1 - \nu  \mathbf{Q}_{K+1},\ \forall 1\leq k \leq K, \label{case2__1} \\
		& \mathbf{F}_k' = \mathbf{F} - \eta_1 \gamma_1 \mathbf{Q}_1 - \nu  \mathbf{Q}_{K+1} ,\ \forall 2\leq k \leq K,\label{case2__2} \\
		&\mathbf{F}_{K+1}= \mathbf{F} - \eta_1 \gamma_1 \mathbf{Q}_1 + \mathbf{C}_{K+1}. \label{case2__3}
	\end{align}
	
	From the equations, we have
	\begin{align}
		&\mathrm{rank}(\mathbf{W}_{c,k}) = 1, 2 \leq k \leq K, \\
		&M_t - 2 \leq \mathrm{rank}(\mathbf{\Phi}_{1}),\mathrm{rank}(\mathbf{\Phi}_{K+1}) \leq M_t - 1.
	\end{align}
	Combining \eqref{case2__2} and \eqref{case2__3}, it follows that 
	\begin{align}
		\mathbf{h}_{K+1}^H \mathbf{e}_{\max}( \mathbf{F} -  \eta_1 \gamma_1 \mathbf{Q}_1 ) = 0,
	\end{align} 
	where $\mathbf{e}_{\max}(\cdot)$ denotes the corresponding eigenvector of the maximum eigenvalue.

Combining \eqref{case2__1} and \eqref{case2__2}, we can obtain that if $\gamma_1 > 1/\eta_1 + 1$, $\mathbf{F}_1'$ only have one maximum eigenvalue, and hence $\mathrm{rank}(\mathbf{W}_{c,1}) = 1$. For the case that $\mathrm{rank}(\mathbf{\Phi}_{K+1}) = M_t - 2$, we can express the transmit beamformers as
	\begin{align}
		&\mathbf{W}_{c,1} = a_1 \mathbf{f}_{\max} \mathbf{f}_{\max}^H + b_1 \mathbf{h}_{1} \mathbf{h}_{1}^H,\\
		&\mathbf{W}_{c,k} = a_k \mathbf{f}_{\max} \mathbf{f}_{\max}^H, 1 \leq k \leq K,\\
		&\mathbf{W}_{c,K+1} = a_{K+1} \mathbf{f}_{\max} \mathbf{f}_{\max}^H +  b_{K+1} \mathbf{h}_{K+1} \mathbf{h}_{K+1}^H.
	\end{align}
	
Similar to case 1, we can construct the optimal solution as 
\begin{align}
	&\mathbf{W}_{c,1}' = (a_1 +  a_{K+1} )\mathbf{f}_{\max} \mathbf{f}_{\max}^H, \\
	&\mathbf{W}_{c,K+1}' =  b_{K+1} \mathbf{h}_{K+1} \mathbf{h}_{K+1}^H,
\end{align}
for the original problem with a rank-one structure and satisfy all the constraints.
It can be readily verified that the constructed solutions are optimal for the original problem and have rank-one property.

	\subsubsection{\textbf{Case 3}} $| \mathcal{K}_1 | \geq 2$. In that case
	\begin{align}
		&\mathbf{\Phi}_{K+1}= \mu \mathbf{I} - \underbrace{(\mathbf{F} - \sum_{k=1}^K \eta_k \Gamma_k \mathbf{Q}_k - \nu \mathbf{Q}_{K+1} )}_{\mathbf{F}'} - \mathbf{C}_{K+1} \\
		&\mathbf{\Phi}_k = \mu \mathbf{I} - \underbrace{(\mathbf{F}- \sum_{i=1}^{K} \eta_i \Gamma_i \mathbf{Q}_i  - \nu  \mathbf{Q}_{K+1}  )}_{\mathbf{F}'}- \eta_k(1 +\Gamma_k ) \mathbf{Q}_k \nonumber \\
		&\qquad = \mu \mathbf{I} - (\mathbf{F}' + \eta_k(1 +\Gamma_k ) \mathbf{Q}_k).
	\end{align}

Note that $\mu \mathbf{I} - \mathbf{F}' \succ \mathbf{0}$ if $\mathbf{D}$ is of full column rank \cite{FanLiu_TSP22_CRB}. However, $\mathbf{\Phi}_k = \mu \mathbf{I} - \mathbf{F}' \succ \mathbf{0}, \forall k \in \mathcal{K}_2$, hence $\mathbf{W}_{c,k} = \mathbf{0}$, which is infeasible. Therefore, we have $| \mathcal{K}_1 | = K$. As a consequence, $\mathbf{\Phi}_k$ is shown as a full-rank positive-definite matrix subtracts a rank-one semidefinite matrix, hence, $\mathrm{rank}(\mathbf{\Phi}_k) = M_t -1, \mathrm{rank}(\mathbf{W}_{c,k}) = 1, \forall 1 \leq k \leq K$.

Given the above discussions on all three cases, it is observed that the optimal solution $\mathbf{W}_{c,k}, \forall 1 \leq k \leq K+1$ always satisfies $\mathrm{rank}(\mathbf{W}_{c,k}) = 1$ if $\mathbf{H} \hat{\mathbf{A}} $ is of full column rank, where $\mathbf{H} = [\mathbf{h}_1,\ldots,\mathbf{h}_{K}, \mathbf{h}_{K+1}]^H \in  \mathbb{C}^{K+1 \times M_t}$, $\hat{\mathbf{A}} = [\mathbf{a}, \dot{\mathbf{a}}]$. This completes the proof.

\balance
\bibliographystyle{IEEEtran}
\bibliography{refs}

\end{document}